\documentstyle[multicol,eqsecnum,aps,prb,epsf]{revtex}
\begin{document}
\title{ Diffusion limited aggregation as a Markovian process. \\
	Part I: bond-sticking conditions}
\author{Boaz Kol and Amnon Aharony}
\address{Raymond and Beverly Sackler Faculty of Exact Sciences, 
	School of Physics and Astronomy,\\ Tel Aviv University, 
	69978 Ramat Aviv, Israel}
\date{\today}
\maketitle
\begin{abstract}
Cylindrical lattice 
Diffusion Limited Aggregation (DLA), with a narrow width $N$, 
is solved using a Markovian matrix method.
This matrix contains the probabilities that the front moves from one
configuration to another at each growth step, calculated
exactly by solving the Laplace equation and using the proper normalization.
The method is applied for a series of approximations, which include
only a finite number of rows near the front.
The matrix is then
used to find the weights of the steady state growing configurations
and the rate of approaching this steady state stage.
The former are then used to find the average upward growth probability, the
average steady-state density and the fractal
dimensionality of the aggregate, which is extrapolated to a value near
1.64.
\end{abstract}
\widetext
\begin{multicols}{2}
\section{Introduction}
Diffusion limited aggregation (DLA) \cite{Witten83} has been the subject of 
extensive study since it was first introduced. This model exhibits a growth
process that produces highly ramified self similar patterns, which are 
believed to be fractals \cite{Mandelbrot82}. It seems that DLA captures the 
essential mechanism in many natural growth processes, such as viscous 
fingering \cite{Feder}, dielectric breakdown \cite{DBM84a}, etc.
It is now understood that the Laplace equation, which is common to
all of these processes and to DLA, has a major role in the resemblance 
between them. One of the interesting features of DLA is that there are no
parameters to fine-tune in order to obtain a fractal. It thus differs from
ordinary critical phenomena, and belongs to the class of self organized
criticality (SOC) \cite{Bak87}. In spite of
the apparent simplicity of the model, an analytic solution is still 
unavailable. Particularly, the exact value of the fractal dimension
is not known.

In DLA there is a seed cluster of particles fixed somewhere. 
A particle is released at a distance from the cluster, and performs a random 
walk until it attempts to penetrate the fixed cluster, in which case it 
sticks. Then the next particle is released and so on. There are two common 
types of sticking conditions. The sticking condition described above
is called ``bond-DLA'', because it occurs when a particle goes into a 
perimeter bond. In ``site-DLA'', sticking occurs as soon as the particle 
arrives in a perimeter site. This paper deals with bond-DLA, whereas part II
deals with site-DLA. The large scale
structure of DLA is not sensitive to the type of sticking conditions used 
\cite{Cafiero93,Erzan95}.

It has been shown that bond-DLA is equivalent to the dielectric breakdown
model (DBM) with $\eta=1$ \cite{DBM84a,DBM84b}. 
DBM is a cellular automaton that is defined on a lattice. It consists of the
following steps: one starts with a seed cluster of connected sites
and with a boundary surface far away from it. 
A field
$\Phi$, which corresponds to the electrostatic potential, is found 
by solving the discrete Laplace equation on a lattice,
\begin{equation}
\nabla^{2} \Phi = 0,
\label{Laplace}
\end{equation}
with the following boundary conditions:
the aggregate is considered to have a constant potential that is 
usually set to $0$, and the potential gradient on the distant boundary 
is set to $1$ in some arbitrary units (some use a constant potential on the 
distant boundary instead). In this paper we set the distant boundary
at infinity, and ignore the exponentially small finite size corrections.
After solving the discrete Laplace equation (\ref{Laplace}), the field $\Phi$
determines the growth probabilities per perimeter bond. More specifically,
the growth probabilities are proportional 
to the electric field to some power $\eta$. The electric field is simply equal
to the  potential difference across each bond. Because the potential
is set to $0$ on the aggregate, the electric-field is equal to 
the potential value at the sites lying across the perimeter 
bonds. Thus,
\begin{equation}
P_{b}={|\Phi_{b}|^{\eta} \over \sum_{b}|\Phi_{b}|^{\eta}}.
\label{growth probabilities}
\end{equation}
Here, $b$ is the bond index.

DLA and DBM can be grown in various geometries.
By geometry we refer to the dimensionality of the lattice,
to the shapes of the boundaries and to the details of the seed for growth 
(usually a point or a line for two dimensional growth). 
For instance, the case in which the distant boundary is 
a sphere is called radial boundary conditions, and
the case in which the boundary is a  distant plane at the top, while the 
seed cluster is a parallel plane at the bottom, with periodic boundary 
conditions on the sides, is called cylindrical boundary conditions.
In this paper we only consider the cylindrical case, with
a relatively short period length (width), from $N=2$ to about $N=7$, 
although the method described here could also be used for wider cases.

\begin{figure}
\epsfysize 9cm
\epsfbox{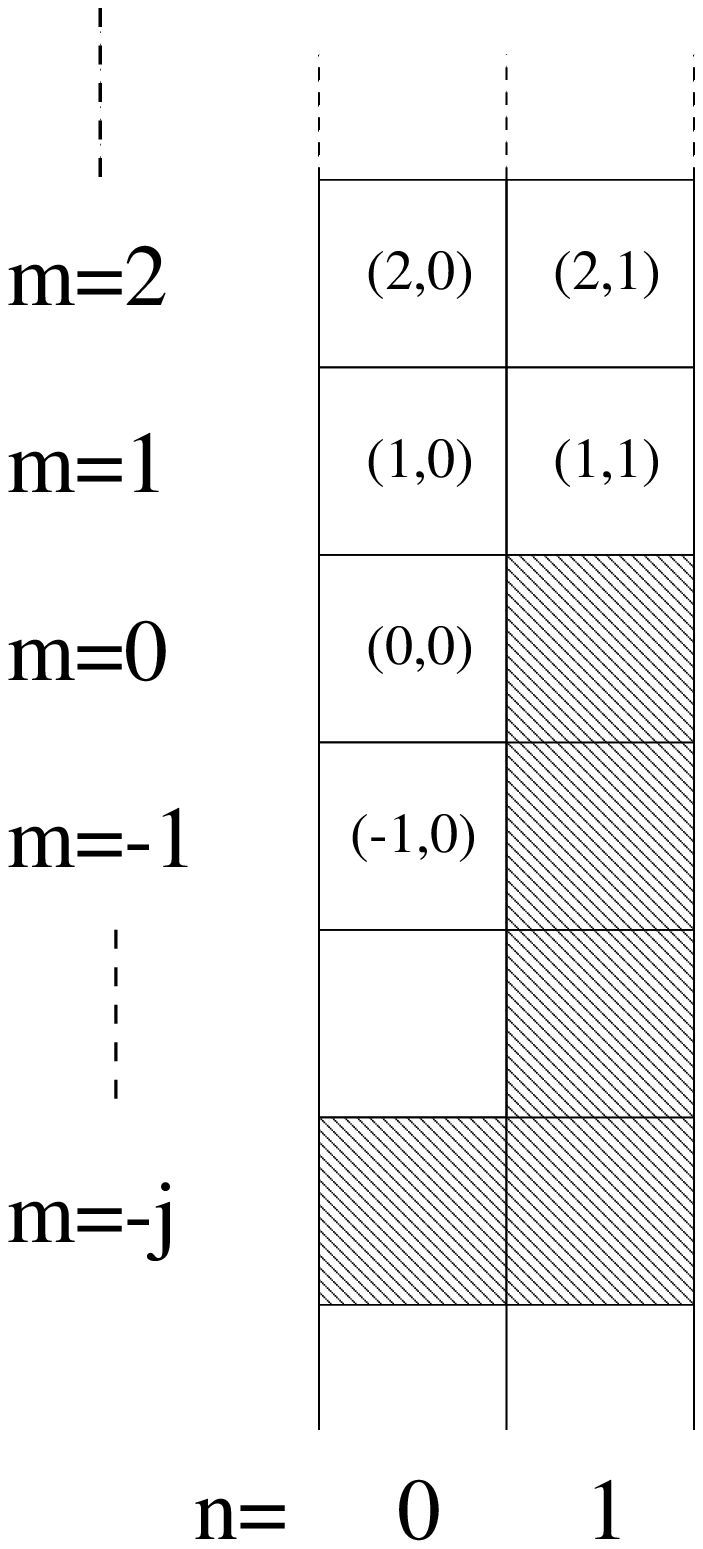}
\vspace{5mm}
\caption{The coordinates $(m,n)$ describe the location on a lattice
that is two sites wide. The gray sites belong to the interface of the 
aggregate, which is shaped as a step of size $j$.}
\label{stepfig}
\end{figure}

\begin{figure}
\epsfysize 9cm
\epsfbox{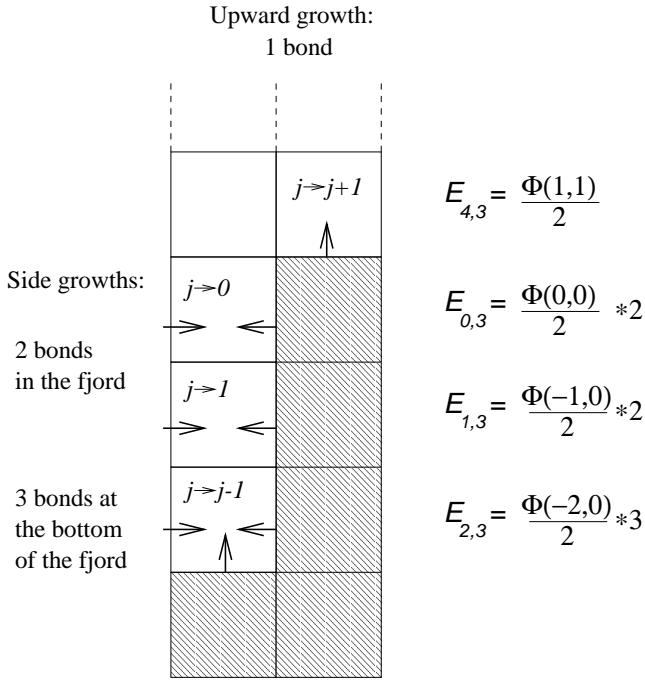}
\vspace{5mm}
\caption{Possible growth processes that change the interface from an initial
step size $j=3$ to a final size $i=4,0,1,2$. 
The growth probability is determined by the potential and the number 
of bonds associated with the site where growth is to occur. $E_{i,j}$ is
the conditional probability to grow from an initial step size $j$ to a final
step size $i$. The normalization follows from Eq. (1.10).}
\label{transitionsfig}
\end{figure}

Recently we published an exact
solution to DLA in cylindrical geometry of width
$N=2$ \cite{Kol98}. The present paper generalizes and extends that solution. 
Our approach follows the dynamics of
the interface. The interface alone determines the growth probabilities at
each time step, and whatever lies behind it is irrelevant. This is because
the solution to the Laplace equation is unique, provided that the
boundary conditions are well defined. We now give a brief summary of Ref. 
\cite{Kol98}. The characterization of the interface
for $N=2$ is simple; The interface is  {\it fully} characterized by a single 
parameter (usually denoted by $i$ or $j$), which corresponds to the height 
difference between the two columns.
This height difference, referred to as the step size, can be infinitely large; 
see Fig. \ref{stepfig}.
If the interface is flat ($j=0$), one can assume that the next particle will
always stick on the right side, without limiting the generality of this 
discussion. This means that the step size can always be considered as 
nonnegative. The Markovian dynamics is then presented using 
the Master equation,
\begin{equation}
P_i(t+1)=\sum_{j=0}^{\infty}E_{i,j}P_j(t),
\label{1.3}
\end{equation}
where $P_j(t)$ is the probability that the step size is $j$ at time $t$,
and  $E_{i,j}$ is the time independent conditional probability that an initial 
step size $j$ will become $i$ after the next growth process. An example with 
several possible transitions is shown in Fig. \ref{transitionsfig}. 
${\bf P}(t)$
is called the state vector and ${\bf E}$ is called the evolution matrix.
In principle, a similar Master equation can be written for more complex
growth situations, provided the various configurations can be indexed
with a single index $j$.
Being made out of conditional probabilities, the elements of the evolution 
matrix obey that, 
\begin{eqnarray}
&0&\leq E_{i,j}\leq 1, ~~ i,j=0,\dots,\infty, \nonumber \\   
&\sum&_{i=0}^{\infty}E_{i,j}=1, ~~j=0,\dots,\infty.
\end{eqnarray}
After many iterations of Eq. (\ref{1.3}) the system converges to a fixed point
${\bf P}^*$, also called the steady state, which represents the asymptotic time
distribution of the step sizes. From the steady state and the evolution matrix 
we are able to extract the average upward growth probability 
$\langle p_{\rm up}\rangle^*$, the average density $\rho$ and the fractal 
dimension $D$.

In order to obtain an analytic expression for the elements of the evolution 
matrix, one  must first solve the Laplace equation. Having found the 
solutions $\Phi(m,n)$, the growth probabilities are found from Eq. 
(\ref{growth probabilities}). The denominator there, which comes 
from the normalization, is particularly simple for the
special case
of $\eta=1$, where the discrete version of the divergence theorem
implies that \cite{Kol98}
\begin{equation}
\sum_b\Phi_b=N.
\label{sum_bPhi_b=N=2}
\end{equation}
The actual growth probability into a site is then found
from
\begin{equation}
p_{\rm site}=\sum_{\rm bonds~into~site}p_b.
\end{equation}

The solution of the Laplace equation is now divided into two parts.
In the
first part, we solve the Laplace equation for the 'upper' part of space,
which starts just above the highest particle of the aggregate and continues
upwards to infinity. In the example of Fig. 1, this part contains all the
rows with $m>0$.
As we explain below, this solution is completely determined by the boundary
conditions and by the values of the potential at the row with $m=0$,
i.e. $\{\Phi(0,n)\}$.
We then solve the Laplace equation for the 'lower' part ($m \le 0$ in
Fig. 1), and find the values of $\{\Phi(0,n)\}$ from matching the two regimes.
The solution in the 'upper' part is given as a combination of solutions
of the form
 \cite{Kol98}
\begin{equation}
\Phi(m,n)=e^{\kappa m +ikn},
\end{equation}
with the dispersion relation
\begin{equation}
\sinh({\kappa \over 2})=\pm \sin({k \over 2})
\label{dispersion}
\end{equation}
and with the discrete set of allowed values $k_l={2\pi \over N}l$, which
follow from the periodic lateral boundary conditions, which require that 
$e^{ikN}=1$.
The boundary conditions at infinity have a uniform gradient, i.e.,
\begin{equation}
\lim_{m \rightarrow \infty} (\Phi(m+1,n)-\Phi(m,n))=1, ~~n=0,\dots,N-1.
\end{equation}
Given the  arbitrarily set of values $\Phi(0,n)$,
the solution for the row $m=1$ is
\begin{equation}
\Phi(1,n)=1+\sum_{n'=0}^{N-1}\Phi(0,n')g_N(|n-n'|),
\label{g}
\end{equation}
where
\begin{equation}
g_N(n)\equiv {1\over N}\sum_{l=0}^{N-1}e^{-\kappa_l}\cos(k_ln),~~n=0,\dots,N-1,
\label{g_N(n)}
\end{equation}
is the boundary Green's function, and $\kappa_l$ corresponds to $k_l$ via the
dispersion relation (\ref{dispersion}). The solution is given only for $m=1$,
because we are only interested in the potential at sites near the interface.
Note that the 
Green's function has the general property  
\begin{equation}
\sum_{n=0}^{N-1}g_N(n)=1
\label{norm}
\end{equation}
\cite{Kol98}. It is therefore good practice to check this normalization
for each of the calculations presented below. Indeed, all our results
obey this rule.

In general, the solution in the 'lower' regime is complicated by the variety of
configurations. However, this solution is very simple for $N=2$, when
$\Phi(m,0)$ is a linear combination of $e^{\kappa_f m}$ and $e^{-\kappa_f m}$.
Since $\Phi(-j,0)=0$, one is left with one unknown $\Phi(0,0)$, to be
determined by the matching at row $0$.

For the special case $N=2$,
the above procedure has led to the exact solution \cite{Kol98}
\begin{eqnarray} 
E_{i,j}&=&\left\{
\begin{array}{cll}
y(\infty)e^{-\kappa_fi}{1-e^{-2\kappa_f(j-i)}\over 1+\beta e^{-2\kappa_fj}}
&,&~0 \leq i \leq j-2 \\
{3\over 2}y(\infty)e^{-\kappa_f(j-1)}{1-e^{-2\kappa_f} \over
1+ \beta e^{-2\kappa_fj}} &,& ~i=j-1 \\
E_{\infty+1,\infty}\left(1-\alpha {e^{-2\kappa_fj} \over 1+\beta 
e^{-2\kappa_fj}} \right) &,& ~i=j+1 \\
0 & & {\rm otherwise}
\end{array}
\right. , \nonumber \\
&&~~j \geq 1,
\label{AnalyticE(i,j)}
\end{eqnarray} 
where 
\begin{equation}
E_{\infty+1,\infty}=\lim_{j\rightarrow \infty}E_{j+1,j}=
{1+g_2(1)y(\infty)\over 2}=0.5658\dots, 
\end{equation}
$y(\infty)=\sqrt{3}-\sqrt{2}=0.3178\dots$, 
$e^{-\kappa_f}=2-\sqrt{3}=0.2679\dots$, 
$\alpha=(1+\beta)g_2(1)y(\infty)/(2E_{\infty+1,\infty})=0.1281\dots$ and
$\beta=5-\sqrt{24}=0.1010\dots$. 
For $j=0$, the interface will
transform into a step of size $j=1$ with probability $1$, hence $E_{1,0}=1$
and $E_{i,0}=0$ for $i \neq 1$.
The values of $E_{i,j}$ for $0 \leq i,j \leq 4$, up to 
the fourth decimal digit, are 
\begin{equation}
{\bf E}=\left[
\begin{array}{cccccc}
0 & 0.4393 & 0.3160 & 0.3177 & 0.3178 & \cdots \\
1 & 0 & 0.1185 & 0.0847 & 0.0851 & \\
0 & 0.5607 & 0 & 0.0318 & 0.0227 & \\
0 & 0 & 0.5655 & 0 & 0.0085 & \\
0 & 0 & 0 & 0.5658 & 0 & \\
\vdots & & & & & \ddots
\end{array}
\right].
\label{ExplicitE(i,j)}
\end{equation}
The first diagonal below the main, 
which represents the probabilities for the step to grow larger by one, 
$E_{j+1,j}$, approaches its asymptotic value of 
$E_{\infty+1,\infty}=0.5658...$ exponentially, as the third row of 
Eq. (\ref{AnalyticE(i,j)}) indicates.
The diagonal above the main represents the probabilities for growths at 
the bottom of the fjord, $E_{j-1,j}$, and corresponds to the second row in
Eq. (\ref{AnalyticE(i,j)}). 
These probabilities decay exponentially as the step size $j$ grows. 
According to the first row in Eq. (\ref{AnalyticE(i,j)}), 
the elements $E_{i,j}$ converge exponentially for
large $j$'s to a simple exponential function:
\begin{equation}
E_{i,\infty}=\lim_{j\rightarrow \infty}E_{i,j}=y(\infty)e^{-\kappa_fi}.
\label{E_(j,infty)}
\end{equation}
These probabilities relate to the transition from a very large step to a 
step of size $i$.  
Next, the steady state vector ${\bf P}^*$ is computed and used 
to evaluate the average upward growth probability 
$\langle p_{\rm up} \rangle^*$,
which in turn, determines the average density $\rho$ and the fractal dimension
$D$. These computations are explained later in Sec. \ref{frustrated}.

Our previous paper does not specify details concerning the manner in which the
system converges to the steady state in time. 
Besides addressing this issue, 
our present paper also treats DLA grown in wider geometrical periods 
(still in cylindrical geometry). 
The basic approach is the same, i.e., we try to characterize the
possible configurations of the interface for wider periods, and then write
the evolution matrix, which is composed of the growth probabilities, which are
computed from the Laplace potential, after proper normalization. 
The first difficulty encountered is in the characterization. For example,
already for a width of $N=3$ one cannot 
characterize the interface using a single
parameter as in the case $N=2$, nor is it easy doing so using $2$ 
parameters, or more. Instead, we make a manual list of possible 
configurations of the interface, which we then
order according to the difference
in height between the highest and lowest points on the interface. 
This difference is denoted by $\Delta m$.
Our 
order-$O$ approximation includes only the configurations with $\Delta m \leq O$.
In our approximation,
some of these configurations represent
many other (excluded) configurations, in the sense that they 
have very similar growth probabilities, especially upward. 
This is because of the screening quality of the Laplace equation, which 
causes the potential to decay exponentially inside fjords. 
Thus, the deeper parts of the interface have a very small effect on the upward
growth probability. 
The finite list of configurations is indexed arbitrarily, with an index 
usually denoted by $i$ or $j$.  
Our experience shows that accurate results are obtained,
only when the order of approximation $O$ is comparable to the width of the
cylinder $N$. Thus, for wide periods, a high order calculation is called for.
This causes the method to be ineffective for very wide periods, because the 
number of configurations grows exponentially with the order of approximation.
We conducted calculation up to $N=7$. 

After selecting the finite list of configurations and obtaining the finite 
evolution matrix,
we compute the steady state vector, which is the fixed point of the matrix
(the normalized eigenvector with an eigenvalue of $1$). For each configuration,
we identify the upward growth processes (when the newly attached particle is 
higher than the rest). We then calculate the average upward growth probability
$\langle p_{\rm up} \rangle^*$ as a weighted average over the configurations.
>From $\langle p_{\rm up} \rangle^*$ we calculate the average density $\rho$ and
the fractal dimension $D$. The computed values of 
$\langle p_{\rm up} \rangle^*$, from different orders of the approximation,
are compared with numerical
simulations in Table \ref{2dresults}.

In Sec. \ref{frustrated} we introduce a simple Markov process, called the
``frustrated climber'', which we solve exactly. A slight modification of the
model is equivalent to site-DLA with a period of $N=2$, which is presented
in part II of this paper \cite{Kol99}.
We then show a way of successively generalizing
the model to approximate bond-DLA with a period of $N=2$ and
with increasing orders $O$.
We are able to
check the approximations by comparing with the exact results of Ref. 
\cite{Kol98}. 
This model also enables us to investigate the
rate of convergence to the steady state. In this context we describe the 
convergence in terms of other eigenvectors, with eigenvalues whose absolute
values are smaller than $1$, and in terms of the infinite shift down operator.
We show that the average upward growth probability converges exponentially in 
time to its steady state value, with a characteristic time constant on
the order of unity. In Sec. \ref{N>2} we generalize our method 
to cylindrical DLA with $N>2$. We present in detail the
calculations for $N=3$ with $O=1$ and $O=2$,
and for 
$N=4$ with $O=1$.
Next we 
report on numerical results for wider periods and higher orders.
In the final section we review the results and summarize.

\section{The frustrated climber model}
\label{frustrated}
Consider someone trying to climb up a slippery infinite ladder. At each time 
step the climber climbs up one step with probability $0\leq p\leq 1$, or falls 
all the way 
down with probability $q\equiv 1-p$. We call the climber ``frustrated'',
because the probability to get very high is exponentially small.
We wish to compute the probability $P_i(t)$ for the climber to be at height 
$i$ after $t$ time steps, for $i=0,\dots,\infty$. 
The Master equation for this problem is ${\bf P}(t+1)={\bf EP}(t)$, where the
matrix element $E_{i,j}$ is the conditional probability that
the climber moves from height $j$ to $i$ in a single
time step. The rules of the model imply that
\begin{equation}
E_{i,j}=\left\{
\begin{array}{ccl}
p&,&~i=j+1 \\
q&,&~i=0 \\
0&,&~\mbox{otherwise}
\end{array}
\right\},~~ j \geq 0,
\end{equation}
so the matrix looks like this:
\begin{equation}
{\bf E}=\left[
\begin{array}{ccccc}
q&q&q&q&\cdots \\ 
p&0&0&0& \\
0&p&0&0& \\
0&0&p&0& \\
\vdots&&&&\ddots
\end{array}
\right].
\label{Eclimb}
\end{equation}
This presentation helps us see the resemblance to the dynamics of DLA with 
$N=2$ in Eqs. (\ref{AnalyticE(i,j)}, \ref{ExplicitE(i,j)}): Eq.
(\ref{Eclimb}) would approximate these equations if we were to replace
$E_{j+1,j}$ by $p \approx E_{\infty+1,\infty}$ 
and $E_{0,j}$ by $q$ for all $j$, 
and neglect all other
growth probabilities, which are indeed smaller. We shall discuss this
and better approximations for DLA in the next subsections.
Because the 
Markovian matrices for the two cases are similar for large $j$'s, we
expect that some of the dynamical features are similar as well. 
We therefore present here an exact solution for the frustrated climber
model, and then try to draw conclusions for generalized models which 
represent successive approximations for DLA.
The advantage is
that in the simple model of the frustrated climber it is possible to derive a
simple analytic expression for the steady state 
and a complete description of the temporal convergence.

The steady state equations for the frustrated climber model are
\begin{eqnarray}
&P^*&_{i+1}=\sum_{j=0}^{\infty}E_{i+1,j}P^*_j=pP^*_i,~~i \geq 0, \\
\Rightarrow &P&^*_j=qp^j,~~j \geq 0.
\end{eqnarray}
One can easily check that this steady state is normalized,
\begin{equation}
\sum_{j=0}^{\infty}P^*_j=\sum_{j=0}^{\infty}qp^j={q\over 1-p}=1.
\end{equation}
The average upward growth probability in the steady state is
\begin{equation}
\langle p_{\rm up} \rangle^* =\sum_{j=0}^{\infty}P^*_jp_{\rm up}(j)
=\sum_{j=0}^{\infty}P_jp=p,
\end{equation}
where $p_{\rm up}(j)$ stands for the probability to move upwards when the 
height of the climber is $j$. In this simple model $p_{\rm up}(j)=p$ for
all $j$'s.

We now investigate the temporal convergence to the steady state. We
define the vector ${\bf v}(t)$ by
\begin{equation}
{\bf P}(t)={\bf P}^*+{\bf v}(t).
\end{equation}
Because ${\bf P}^*$ and ${\bf P}(t)$ are probability vectors,
$\sum_{j=0}^{\infty}P^*_j=\sum_{j=0}^{\infty}P_j(t)=1$, for any $t$, hence 
\begin{equation}
\sum_{j=0}^{\infty}v_j(t)=0.
\label{sum_v=0}
\end{equation}
We substitute ${\bf v}$ into the dynamical equation and obtain
\begin{eqnarray}
&{\bf P}&(t+1)={\bf EP}(t)={\bf P}^*+{\bf Ev}(t), \\
\Rightarrow &{\bf v}&(t+1)={\bf Ev}(t).
\end{eqnarray}
Next, we look for
the rest of the eigenvectors of the evolution matrix
(any eigenvector ${\bf v}$ with an eigenvalue 
$\lambda \neq 1$, has to obey Eq. (\ref{sum_v=0})). Surprisingly, 
there are no eigenvectors besides the steady state in this case.
The eigenvector equations are
\begin{eqnarray}
&&\lambda v_0=q\sum_{j=0}^{\infty}v_j=0, \nonumber \\
&&\lambda v_{i+1}=pv_i(t),~~ i \geq 0.
\label{other_eigenvalues}
\end{eqnarray}
The first equation implies that either $\lambda=0$ or $v_0=0$. In both 
cases, the last equation implies that {\bf v}=0.

We next introduce the infinite shift-down operator:
\begin{equation}
{\bf S} \equiv \left[
\begin{array}{ccccc}
0&0&0&0&\cdots \\ 
1&0&0&0& \\
0&1&0&0& \\
0&0&1&0& \\
\vdots&&&&\ddots
\end{array}
\right].
\end{equation}
This operator causes a vector to ``slide down'' and inserts a
zero at the evacuated component at the top. {\bf S} has no eigenvectors at all,
not even a fixed point (in spite of the fact that $\sum_{i=0}^{\infty}S_{i,j}
=1$ for $j=0,\dots,\infty$).
In fact, ${\bf Ev}=p{\bf Sv}$ for all vectors ${\bf v}$ with
$\sum_{j=0}^{\infty}v_j=0$.

Nevertheless, the convergence of ${\bf P}(t)$ to ${\bf P}^*$ is simple. 
Starting from any initial state vector ${\bf P}(t=0)$, the first application
of ${\bf E}$ causes the first component to be set to its steady state 
value $P_0(t=1)=q$. 
At each subsequent iteration another components is set permanently: 
$P_1(t=2)=qp$, $P_2(t=3)=qp^2$, etc. $P_j$ becomes equal to $P^*_j$ 
after no more than $j+1$ time steps. 
The context we are interested in is wider. We wish to
compute the convergence of ``observables'', i.e., the average
of an arbitrary function $a(j)$ over configurations. 
We compute the average at time $t$
\begin{equation}
\langle a \rangle(t) \equiv \sum_{j=0}^{\infty}a(j)P_j(t)
=\langle a \rangle ^* +\sum_{j=0}^{\infty}a(j)v_j(t),
\end{equation}
where $\langle a \rangle^* \equiv \sum_{j=0}^{\infty}a(j)P_j^*$ is the steady
state average. 
Starting from an initial deviation from the steady state ${\bf v}(0)$, each
iteration causes a down shift and a multiplication by $p$, hence
\begin{equation}
\langle a \rangle(t)=\langle a \rangle ^*+p^t\sum_{j=0}^{\infty}a(j+t)v_j(0).
\label{converge}
\end{equation}
Equation (\ref{converge}) is the analogue of the standard eigenvector 
description. We can also identify here the exponential decay of the factor 
$p^t$.
For example, the function $a(j)=\delta_{j,j_0}$ ``measures'' the probability
of the climber to be at height $j_0$ (at any time). At time $t$ the observed
average probability is
\begin{equation}
\langle a \rangle (t)=P^*_{j_0}+p^tv_{j_0-t}(0),
\end{equation}
for $t \leq j_0$, and $\langle a \rangle (t)=P^*_{j_0}$ for $t>j_0$ 
\cite{remark}.
\subsection{First-order approximation for $N=2$}
\label{site stick}
We now return to Eq. (\ref{AnalyticE(i,j)}), and try to approximate it
by a sequence of models which are related to the frustrated climber model.
The simplest approximation would follow if we do not let
the particle penetrate into the fjord at all. This is equivalent to setting
$\kappa_f = \infty$ in Eq. (\ref{AnalyticE(i,j)}). According to these simplified
rules, the particle can either stick at $(0,0)$
and create a flat step of $i=0$, or it can stick at $(1,1)$ and increase the
step height by $1$. 
Let us denote the probability for the
former event by $q$ and the latter by $p$.
In the first-order approximation we take $p$ and $q$ to be
independent of the initial step size $j$, unless $j=0$, in which case
the step size increases with probability $1$. 
The Markovian matrix {\bf E} for
this case is almost identical to the case of the frustrated climber,
\begin{equation}
{\bf E}=\left[
\begin{array}{ccccc}
q_0&q&q&q&\cdots \\ 
p_0&0&0&0& \\
0&p&0&0& \\
0&0&p&0& \\
\vdots&&&&\ddots
\end{array}
\right],
\label{q_0matrix}
\end{equation}
the only difference being in the first column, where we denote $q_0=0$ and
$p_0=1$. In part II of this paper we show that this model is exact for 
the case of site-sticking DLA for $N=2$
\cite{Kol99}.

The solution to this problem is very similar to that of the frustrated climber,
with small modifications. The steady state is
\begin{equation}
P^*_j=P^*_0p_0p^{j-1},~~j \geq 1,
\end{equation}
where $P^*_0$ can be determined using the normalization condition
\begin{eqnarray}
&&1=\sum_{j=0}^{\infty}P^*_j=P_0^*(1+p_0\sum_{j=0}^{\infty}p^j), \nonumber \\
&&\Rightarrow P^*_0={1-p \over 1-p+p_0}.
\end{eqnarray}
The average upward growth probability is evaluated by
\begin{equation}
\langle p_{\rm up}^{(1)}\rangle^*=P^*_0p_0+(1-P^*_0)p={p_0 
\over 1-p+p_0}.
\label{pup(1)}
\end{equation}
The superscript $(1)$ appears because it is the first-order approximation.
We now need to choose $p$. One possible choice would be
to take $p=E_{\infty+1,\infty}=0.5658$, because this is the asymptotic
upward growth probability, and then set $q=1-p$.
This would give $\langle p_{\rm up}^{(1)}\rangle^*=0.6973$,
to be compared with the exact value $0.6812$ \cite{Kol98}. 
An alternative approximation would return to Eq. (1.13), but 
replace $y(\infty)$ by $q$, and then find $q$ by solving
$1=p+q=[1+g_2(1)q]/2+q$. This yields $p=1-q=2-\sqrt{2}=0.5858$, and therefore
$\langle p_{\rm up}^{(1)}\rangle^*=\sqrt{2}/2=0.7071$.

We next calculate the average density and the fractal dimensionality.
Similar to the argument used by Turkevich and Scher \cite{Turkevich89}, we
consider a large number of growth processes $n$ in the steady state. 
During this growth the aggregate would grow higher by $h=\langle p_{\rm up}
\rangle^* n$. 
The total volume  covered by the new growth is $hN^{d-1}$, where $d=2$ is the 
Euclidean dimension. Thus, for $N=2$ and for our first approximation
the density is
\begin{equation}
\rho={n \over hN^{d-1}}={n \over \langle p_{\rm up} \rangle^* nN^{d-1}}
={1 \over \langle p_{\rm up}\rangle^* N^{d-1}}=0.7171,
\label{rho}
\end{equation}
to be compared with the exact value $\rho=0.7340$.
Although our model does not really produce fractal structures (due to the 
narrow width of our space), we can make an estimate of the fractal dimension
in the same way Pietronero $et~al.$ estimated it in 
\cite{Erzan95,Pietronero88b}. For a self similar fractal
structure, one expects that a change of scale by a factor $N$ will change the
average mass (number of occupied sites) of a $N \times N$ cut by a factor 
$N^D$, where $D$ is the fractal dimension. Assuming that the above procedure
represents a coarse graining of the sites into $N \times N$ cells, we conclude
that asymptotically 
\begin{equation}
\rho =N^{D-d},
\label{rho=N^(D-d)} 
\end{equation}
and this means that
\begin{equation}
D=d+{\ln(\rho) \over \ln(N)}=1-{\ln(\langle p_{\rm up} \rangle^*) \over
\ln(N)}=1.5202. 
\label{estimate dimension}
\end{equation}
In Sec. IV we suggest a modified estimate of the fractal dimension, allowing
for corrections to the asymptotic form (\ref{rho=N^(D-d)}). 

The study of the convergence to the steady state is again limited to the 
subspace of vectors ${\bf v}$ with $\sum_{j=0}^{\infty}v_j=0$. 
The dynamic equation for $i=0$ is,
\begin{eqnarray}
&v&_0(t+1)=q_0v_0(t)+\sum_{j=1}^{\infty}qv_j(t)=(q_0-q)v_0(t), \nonumber \\
\Rightarrow &v&_0(t)=(q_0-q)^tv_0(0).
\end{eqnarray}
Since $q_0=0$, the exponentiated prefactor is negative, and therefore $v_0(t)$
is oscillating during its decay.
After the first iteration $v_1(1)=p_0v_0(0)$, regardless of its initial value.
Afterwards it continues to follow $v_0$, i.e., $v_1(t)=p_0(q_0-q)^{t-1}v_0(0)$.
After the second iteration $v_2(2)=p_0pv_0(0)$, and it also starts to decay
exponentially with the factor $(q_0-q)$. This happens for any $j>1$; 
After more than $j$ time steps ($t>j$) one has,
\begin{equation}
v_j(t)=p_0p^{j-1}(q_0-q)^{t-j}v_0(0).
\end{equation}
For short times and large indices $t<j$, the dynamics is governed by the shift
down operator:
\begin{equation}
{\bf v}(t)=v_0(0)(q_0-q)^t{\bf h}+p^t\sum_{j=1}^{\infty}c_j
{\bf e}^{(j+t)},
\label{v(t)(1)}
\end{equation}
where ${\bf e}^{(j)}$ are the standard basis vectors, the components
of the vector ${\bf h}$ are, 
\begin{eqnarray}
h_0&\equiv& 1, \nonumber \\
h_j&\equiv& {p_0 \over p} \left( {p \over q_0-q}\right)^j,~~j \geq 1, 
\end{eqnarray}
and the constants $c_j$ are determined by the initial conditions,
\begin{equation}
c_j=v_j(0)-v_0(0)h_j, ~~j=1,2,\dots
\end{equation}
For $p>0.5$ the components of ${\bf h}$ explode 
exponentially. However, $\sum_{j=0}^{\infty}v_j(0)=0$ and therefore 
$\lim_{j\to \infty}v_j(0)=0$. Thus, in order to cancel the divergence of the
$h_j$'s, the $c_j$'s must also explode exponentially and have an opposite
sign. We note that because of this divergence ${\bf h}$
does not have a finite $L_1$ norm and thus
does not belong to the domain of ${\bf E}$. Therefore it is not
an eigenvector.

\subsection{Higher-order approximations for $N=2$}
\label{approxN=2}
As mentioned earlier, the frustrated climber model resembles the bond-DLA 
evolution matrix (\ref{AnalyticE(i,j)}, \ref{ExplicitE(i,j)}).
In this section we approximate the full
dynamics using increasingly more complex matrices.
By doing so we do not improve on the accuracy of our previously published
results \cite{Kol98}, but rather learn about the rate of convergence to the
steady state. The method used in this section is generalized and applied
to cylindrical DLA with wider periods in the next section. 
The case $N=2$ is the simplest demonstration of this approach.

The second-order approximation is to allow also transitions of the kind
$j \rightarrow 1$ for $j \geq 1$. 
We also allow having arbitrary values in
the top left $2 \times 2$ corner of the matrix, which we copy from
the original matrix of Eq. (\ref{ExplicitE(i,j)}), i.e.,
\begin{equation}
{\bf E}=\left[
\begin{array}{cccccc}
q_0&q_1&q&q&q&\cdots \\ 
r_0&r_1&r&r&r& \\
0&p_1&0&0&0& \\
0&0&p&0&0& \\
0&0&0&p&0& \\
\vdots&&&&&\ddots
\end{array}
\right],
\label{q_1matrix}
\end{equation}
We still require that the sum of the elements in each column be 
equal to $1$, i.e.,
\begin{eqnarray}
&&q_0+r_0=1, \nonumber \\
&&q_1+r_1+p_1=1, \nonumber \\
&&q+r+p=1.
\end{eqnarray}
In terms of standard DLA this means that
we allow the particle to penetrate two sites into the fjord, but no more.
Indeed it is exponentially improbable to penetrate deep into the fjord. This 
fact suggests a controlled approximation for DLA. In each order of the 
approximation we allow the depth of penetration into the fjord to grow by $1$.
This is done by copying the $(O+1)\times O$ upper left block of the original 
matrix (\ref{AnalyticE(i,j)}, \ref{ExplicitE(i,j)}), where $O$ is the order of 
approximation. Asymptotic values are used outside this block, i.e.,
\begin{eqnarray}
&&E_{j+1,j}=E_{\infty+1,\infty},~~ j \geq O, \nonumber \\
&&E_{i,j}=y(\infty)e^{-\kappa_fi},~~ j \geq O,~ i \leq O-2, \nonumber \\
&&E_{n-1,j}=1-\sum_{i=0}^{n-2}y(\infty)e^{-\kappa_fi}-E_{\infty+1,\infty}
\nonumber \\
&&~~=y(\infty){e^{-\kappa_f(n-1)} \over 1-e^{-\kappa_f}}, ~~ j \geq O,
\end{eqnarray} 
and the rest of the matrix elements are null.
For example, in our case, $O=2$, the constants in the matrix (\ref{q_1matrix})
are
\begin{eqnarray}
&&q_0=0, \nonumber \\
&&r_0=1, \nonumber \\
&&q_1={6-3\sqrt{2} \over 4}=0.4393, \nonumber \\
&&r_1=0, \nonumber \\
&&p_1={3\sqrt{2}-2 \over 4}=0.5607, \nonumber \\
&&q=y(\infty)=\sqrt{3}-\sqrt{2}=0.3178, \nonumber \\
&&p=E_{\infty+1,\infty}=0.5658, \nonumber \\
&&r=y(\infty){e^{-\kappa_f} \over 1-e^{-\kappa_f}}=0.1163.
\label{q_1constants}
\end{eqnarray}

First, the steady state is found by solving ${\bf P}^*={\bf EP}^*$, i.e.,
\begin{eqnarray}
&&P^*_0=q_0P^*_0+q_1P^*_1+q\sum_{j=2}^{\infty}P^*_j, \nonumber \\
&&P^*_1=q_0P^*_0+q_1P^*_1+r\sum_{j=2}^{\infty}P^*_j, \nonumber \\
&&P^*_2=p_1P^*_1, \nonumber \\
&&P^*_{j+1}=pP^*_j,~~ j \geq 2.
\label{q_1steady}
\end{eqnarray}
The solution to the last equation is
\begin{equation}
P^*_j=P^*_2p^{j-2}, ~~ j \geq 2.
\end{equation}
Keeping this in mind it is possible to exchange the two last equations of the
set (\ref{q_1steady}) with
\begin{equation}
\sum_{j=2}^{\infty}P^*_j=p_1P^*_1+p\sum_{j=2}^{\infty}P^*_j.
\end{equation}
Thus we obtain an autonomous and finite set of $3$ equations for $3$ unknowns,
namely, $P^*_0$, $P^*_1$ and $\tilde P^*_2 \equiv \sum_{j=2}^{\infty}P^*_j$. 
The third parameter, $\tilde P^*_2$, represents the total probability
for the infinitely many
 configurations with
$j\geq 2$. This reduction of the problem to three parameters
became possible because all of
the configurations with $j\geq 2$
have exactly the same transition probabilities to
the configurations $j=0$ and $j=1$, and because they have exactly the same
upward growth probability. 
Thus we obtain a fixed point equation for a $3\times 3$ matrix,
\begin{equation}
\left[
\begin{array}{c}
P^*_0 \\
P^*_1 \\
\tilde P^*_2 \\
\end{array}
\right]=\left[
\begin{array}{ccc}
q_0 & q_1 & q \\
r_0 & r_1 & r \\
0 & p_1 & p \\
\end{array}
\right] \left[
\begin{array}{c}
P^*_0 \\
P^*_1 \\
\tilde P^*_2 \\
\end{array}
\right].
\label{q_1finite}
\end{equation}
It is guaranteed that a nontrivial solution exists, because the sum of 
the terms in each 
column of the finite matrix equals $1$. Using the constants from
Eqs. (\ref{q_1constants}), the normalized solution obtained is,
\begin{eqnarray}
&&P_0^{*(2)}=0.2705,~~(0.2696), \nonumber \\
&&P_1^{*(2)}=0.3184,~~(0.3113), \nonumber \\
&&\tilde P_2^{*(2)}=0.4111,~~ (0.4191), 
\end{eqnarray}
where the superscript denotes the order of approximation and a comparison is
drawn to the exact values in parentheses. By ``exact'' we refer to very high
order calculations, or to values from simulations (which are the same up to 
the presented accuracy of $10^{-4}$) \cite{Kol98}. The elements 
$P^*_j$ for $j \geq 2$ are evaluated using
\begin{equation}
P_j^{*(2)}=(1-p)\tilde P_2^{*(2)}p^{j-2}, ~~j \geq 2.
\end{equation}
It is now possible to evaluate the average upward growth probability
\begin{equation}
\langle p_{\rm up}^{(2)} \rangle^*=P^*_0r_0+P^*_1p_1+\tilde P^*_2p=0.6816,
\end{equation}
where the exact value is $0.6812$. The fractal dimension is evaluated 
as in Eq. (\ref{estimate dimension}),
\begin{equation}
D^{(2)}=1.5530,
\end{equation}
compared to the exact value $1.5538$. 

The temporal convergence to the steady state in the second-order approximation
can be analyzed using both the shift down operator and eigenvectors.
The first eigenvector of the matrix in 
Eq. (\ref{q_1finite}) is the fixed point solution, 
which we denote by $\tilde {\bf P}^*$.
Let us denote the other two (three-components) eigenvectors 
by $\tilde {\bf h}$ and $\tilde {\bf g}$, 
and their corresponding eigenvalues by $|\lambda_0|\geq|\lambda_1|$.
After $t$ iterations of the evolution matrix we have
\begin{equation}
\tilde {\bf P}(t)=\tilde {\bf P}^*+c_0\lambda_0^t\tilde {\bf h}
+c_1\lambda_1^t\tilde {\bf g},
\label{finite eigenvectors}
\end{equation}
where $c_0$ and $c_1$ are constants determined by the initial conditions.
The configurational average of some function $a(j)$ with $a(j)=a(2)$
for $j>2$, can be expressed in terms of these eigenvalues only,
\begin{equation}
\langle a \rangle(t)=\langle a \rangle ^*+k_0\lambda_0^t +k_1\lambda_1^t,
\end{equation}
where $k_0$ and $k_1$ are some other constants. 
A special function of this type is the upward growth probability, 
$p_{\rm up}(j)=(r_0, p_1, p, p, p,\dots)$. 
The eigenvalue with the largest absolute
value other than $1$, $\lambda_0$, makes the largest contribution to the
deviation from the steady state values, and thus controls the temporal 
convergence. The characteristic time constant for the exponential convergence
is,
\begin{equation}
\tau=-{1 \over \ln(|\lambda_0|)}.
\label{tau=}
\end{equation}
The eigenvalues obtained are $\lambda_0^{(2)}=-0.5599$ and 
$\lambda_1^{(2)}=0.1257$, using the constants of Eqs. 
(\ref{q_1constants}). Hence, $\tau^{(2)}=1.7$.
In order to describe the convergence of $P_j(t)$ for $j \geq 2$ we use 
the vector ${\bf v}(t)=
{\bf P}(t)-{\bf P}^*$, once more, and we perform a decomposition similar to 
Eq. (\ref{v(t)(1)}):
\begin{equation}
{\bf v}(t)=c_0\lambda_0^t{\bf h}+c_1\lambda_1^t{\bf g}+p^t\sum_{j=2}^{\infty}
c_j{\bf e}^{(j+t)}, 
\label{v(t)(2)}
\end{equation}
where $c_0$ and $c_1$ are the same as in Eq. (\ref{finite eigenvectors}) and
the constants $c_j$ for $j \geq 2$ are determined by the initial condition
${\bf v}(0)$. The vectors ${\bf h}$ and ${\bf g}$ are infinite 
generalizations of the finite vectors $\tilde {\bf h}$ and $\tilde {\bf g}$,
according to
\begin{equation}
\begin{array}{lll}
h_j=\tilde h_j,&g_j=\tilde g_j, & j=0,1, \\
h_2=p_1\tilde h_1,&g_2=p_1\tilde g_1, &j=2,\\
h_j=h_2\left({p\over \lambda_0} \right)^{j-2},
&g_j=g_2\left({p\over \lambda_1} \right)^{j-2},& j\geq 2.
\end{array}
\end{equation}
Because $p=E_{\infty+1,\infty}>|\lambda_0|,|\lambda_1|$, it is apparent that
the components $h_j$ and $g_j$ diverge exponentially for large $j$'s. This
means that these vectors do not have a finite $L_1$ norm, and that they do
not belong to the domain of ${\bf E}$. Therefore, they are not eigenvectors,
and $\lambda_0$ and $\lambda_1$ are not eigenvalues of ${\bf E}$. Nevertheless,
Eq. (\ref{v(t)(2)}) is still true. The effect of the shift down operator
is manifested in the sum $p^t\sum_{j=2}^{\infty}c_j{\bf e}^{(j+t)}$. 

Using the same method it is possible to make higher order calculations. The 
steady state quantities resulting from the third order approximation are
presented in Table \ref{steady3}, in comparison with exact results.
The eigenvalue with the largest absolute value is $\lambda_0^{(3)}=-0.5687$,
which has a greater absolute value than $E_{\infty+1,\infty}=0.5658$. This
means that a legitimate eigenvector exists for the infinite matrix.
In the fourth and fifth order approximation we get 
$\lambda_0^{(4,5)}\approx -0.5688$.
This suggests that the higher the order the more accurate is the evaluation of
$\lambda_0$ and that the accuracy obtained is better than $10^{-4}$. The 
typical time needed to settle in the steady state from any initial condition 
is therefore as short as 
\begin{equation}
\tau=1.8.
\end{equation}

\section{DLA with $N>2$}
\label{N>2}
\begin{figure}
\epsfbox{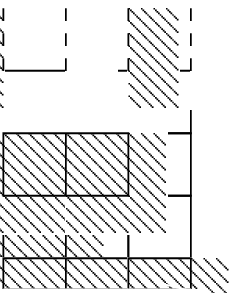}
\vspace{5mm}
\caption{An example of an interface configuration for $N=3$ that cannot be
characterized using the height differences of two columns relative to the 
third.}
\label{N3examplefig}
\end{figure}

\begin{figure}
\epsfbox{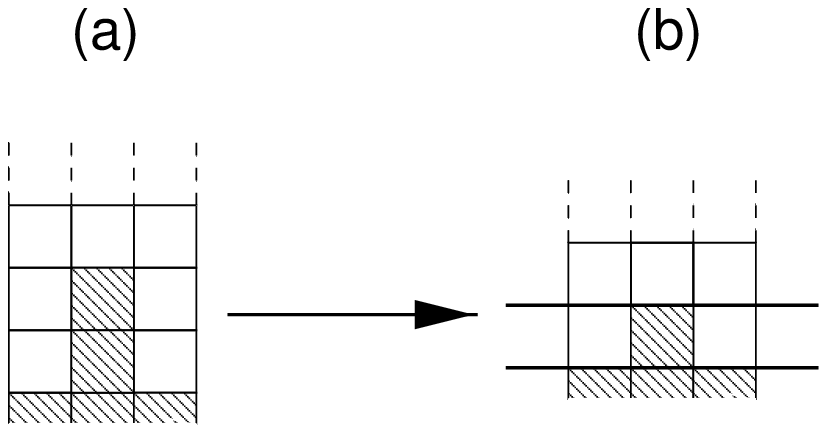}
\vspace{5mm}
\caption{Configuration (a), with $\Delta m=2$, is truncated by 
taking only the top row, and turns into configuration (b), with 
$\Delta m=1$, in the first-order approximation ($O=1$).}
\label{truncatefig}
\end{figure}

The generalization of the exact methods from Ref.
\cite{Kol98} to $N>2$ is
not straightforward. Trying to proceed along a similar line, one would try to 
parameterize the interface with a parameter $i=1,2,\dots,\infty$, and write
the Master equation $P_i(t+1)=\sum_{j=1}^{\infty}E_{i,j}P_j(t)$. Unlike the 
case $N=2$, the parameterization for $N>2$ is very complicated. 
For instance, for the case $N=3$ it is reasonable to try
using two parameters, which indicate the height of two columns relative to
the highest (or lowest) third column. However,
this is insufficient because complex
fjords (involving overhangs) might occur, as shown in Fig. \ref{N3examplefig}.
Instead of achieving
a perfect parameterization, we adopt the approximate approach of 
Sec. \ref{approxN=2},
i.e., we take into account only a finite number of interface configurations. 
These configurations are classified 
according to the maximum height difference between the highest and lowest 
particles on the interface $\Delta m$. In the $O$th-order 
approximation all the configurations with $\Delta m \leq O$ are 
included. The excluded configurations with $\Delta m> O$ are 
transformed into a configuration with $\Delta m=O$, by filling in
the $(O+1)$th row below the highest particle; see Fig. \ref{truncatefig}. 
This transformation does not
change the growth probabilities considerably. Especially, the upward growth 
probability would hardly change for large $O$. 
The variable $P_i(t)$, where $i$ 
corresponds to a configuration with $\Delta m=O$, actually represents
the sum of probabilities of all the configurations with 
$\Delta m \geq O$, that have the same $O$ uppermost rows, 
rather than represent the probability of the configuration $i$ alone.
This is analogous to $\tilde P_2^*$ in the example above, see 
Sec. \ref{approxN=2}. 
After the finite set of configurations
is chosen, the configurations are indexed with arbitrary 
consecutive numbers. Then, the growth probabilities for each configuration are
computed by solving the Laplace equation and by taking
into account the bond multiplicity. Each growth process results in a different
final configuration, which must be identified with one of the configurations
in the finite set. Special attention is required for the upward growth 
processes, which might result in configurations with $\Delta m>O$,
which do not belong to the finite set. This is rectified by truncating the
bottom row of the interface (considering it as fully occupied). 
The total upward probability for each 
configuration is added up and stored in a function $p_{\rm up}(i)$, later
to be averaged over the steady state distribution of configurations. The
growth probabilities are arranged in the evolution matrix, ${\bf E}$, whose
fixed point corresponds to the steady state distribution of configurations, 
which is required for evaluating $\langle p_{\rm up} \rangle^*$, $\rho$ 
and $D$. Because the matrix
is finite, the existence of at least one fixed point is guaranteed. The other 
eigenvectors describe the rate of convergence to the steady state. 

The best way to demonstrate this approach is by showing a few
sample calculations. The easiest ones are the first and second order 
approximation for $N=3$ and the first order approximation for $N=4$.
After that we explain the general algorithm for higher orders and widths,
and report the results obtained numerically.
\subsection{First order approximation for $N=3$}
\begin{figure}
\epsfbox{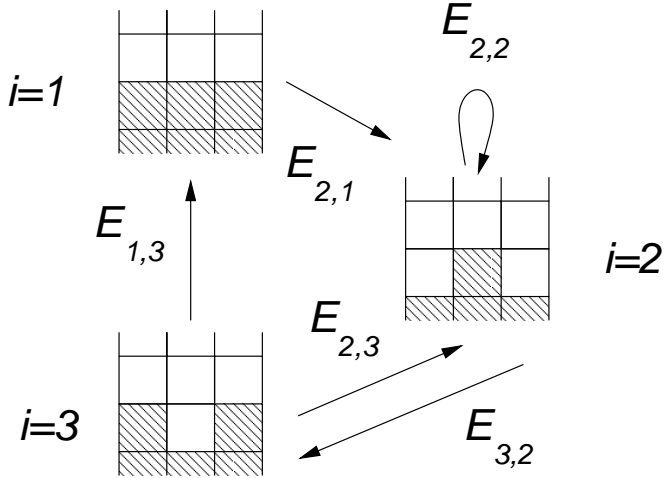}
\vspace{5mm}
\caption{The three possible configurations in the first-order 
approximation for $N=3$, up to translation symmetry. 
The arrows indicate the possible 
transitions due to growth processes. The transition probability from
configuration $j$ to $i$ is denoted by $E_{i,j}$.}
\label{N3o1indexfig}
\end{figure}

In the first order approximation we only look at the top row of the
aggregate. For $N=3$ there are only $3$ possible configurations (up to
symmetry), with the top row occupied by $1$, $2$ or $3$ particles. 
Each configuration is indexed and for each configuration we identify the 
growth processes and the final configurations resulting from them; see
Fig. \ref{N3o1indexfig}. In part II of this paper we show that the calculation
presented in this section can be used to solve exactly (no approximations)
the case of site-DLA with $N=3$.

The first configuration $(j=1)$ grows upward with probability $1$,
thus $p_{\rm up}(1)=1$. The resulting configuration is $i=2$, thus $E_{2,1}=1$
and $E_{i,1}=0$ for $i \neq 2$. This concludes the construction of the
first column of the evolution matrix. 

In order to obtain the other growth
probabilities we have to solve the relevant Laplace problems, for which we need
the Green's function according to Eq. (\ref{g_N(n)}). For $N=3$ we have 
$k_l={2\pi \over 3}l$ for $l=0,1,2$. We recall that $e^{-\kappa (k)}=
q-\sqrt{q^2-1}$, where $q \equiv 2-\cos(k)$ \cite{Kol98} and find that
\begin{eqnarray}
&&e^{-\kappa_0}=1, \nonumber \\
&&e^{-\kappa_1}=e^{-\kappa_2}={5-\sqrt{21}\over 2},
\end{eqnarray}
and thus
\begin{eqnarray}
g_3(0)={1 \over 3}\left(1+2{5-\sqrt{21}\over 2}\right)={6-\sqrt{21}\over 3},
\nonumber \\
g_3(1)=g_3(2)={1-g_3(0) \over 2}={\sqrt{21}-3\over 6}.
\end{eqnarray}
These values obey the normalization condition (\ref{norm}).

\begin{figure}
\epsfbox{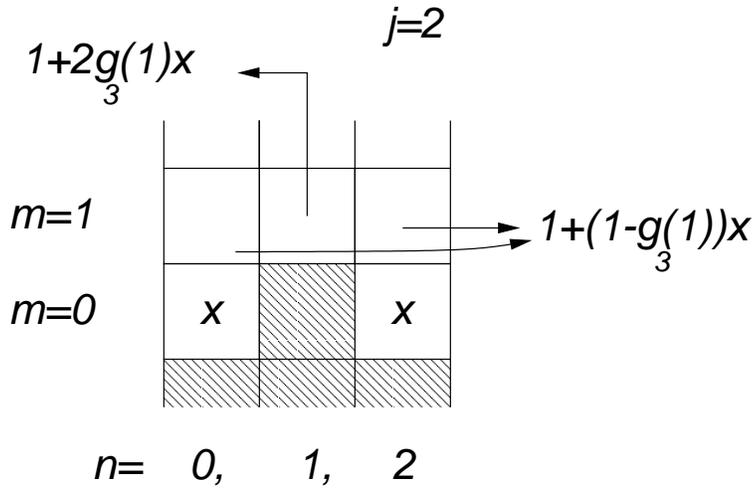}
\vspace{5mm}
\caption{A ``potential diagram'': the potentials $\Phi(m,n)$ 
of the configuration $j=2$, expressed in terms of the variable $x$.}
\label{N3o1j=2}
\end{figure}

Because of the symmetry of the configuration $j=2$, the potential can be 
expressed in terms of one variable $x\equiv \Phi(0,0)=\Phi(0,2)$, as shown
in Fig. \ref{N3o1j=2}. This kind of figure demonstrates the distribution of 
the potential $\Phi(m,n)$ over the lattice, and thus we call it a ``potential
diagram''. The potentials $\Phi(1,0)=\Phi(1,2)=1+(1-g_3(1))x$ do
not correspond to a growth process, but are important for solving for $x$. The
potential $\Phi(1,1)=1+2xg_3(1)$ corresponds to the upward growth process.
The Laplace equation for $x$ is
\begin{eqnarray}
4x&=&x+(1-g_3(1))x+1, \nonumber \\
\Rightarrow x&=&{9-\sqrt{21}\over 10}=0.4417.
\end{eqnarray}
Growth in both sites $(0,0)$ and $(0,2)$ results in configuration $i=3$,
hence 
\begin{equation}
E_{3,2}={4 \over 3}x={18-2\sqrt{21} \over 15}=0.5890,
\end{equation}
where the numerator, $4$, is inserted because there are $2$ 
bonds for each of the $2$ growth sites, and the denominator is the 
normalization factor $N=3$. A growth process in site $(1,1)$ results in an
interface that does not belong to our finite set. In this approximation we
only take into account the top most row of the interface, and therefore this
interface is identified with configuration $i=2$, i.e., 
\begin{equation}
E_{2,2}={2xg_3(1)+1\over 3}={2\sqrt{21}-3\over 15}=0.4110.
\end{equation} 
The transition to $i=1$ is impossible, hence, $E_{1,2}=0$. It is easy to check
that
the second column of the matrix is normalized, i.e., $\sum_{i=1}^3E_{i,2}=1$. 
The total upward growth probability for this configuration is 
\begin{equation}
p_{\rm up}(2)=E_{2,2}=0.4110.
\end{equation}

\begin{figure}
\epsfbox{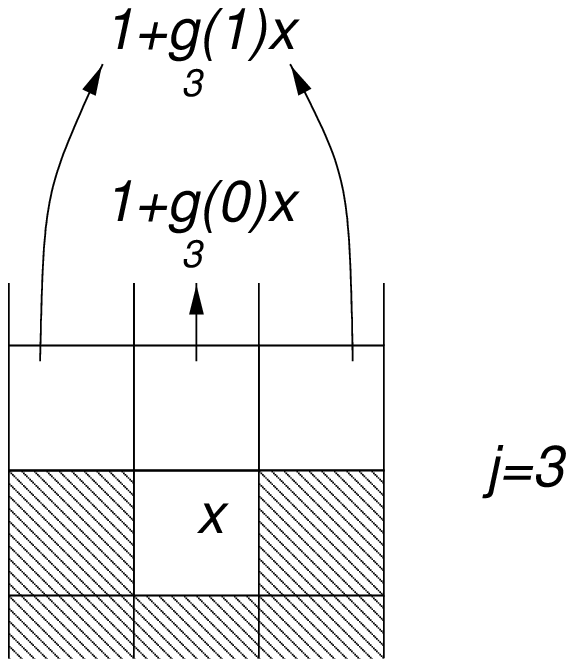}
\vspace{5mm}
\caption{The ``potential diagram'' for configuration $j=3$.}
\label{N3o1j=3}
\end{figure}

The potentials of configuration $j=3$ are described in terms of $x=\Phi(0,1)$,
as in Fig. \ref{N3o1j=3}. The Laplace equation is 
\begin{eqnarray}
4x&=&g_3(0)x+1, \nonumber \\
\Rightarrow x&=&{6-\sqrt{21}\over 5}=0.2835.
\end{eqnarray}
There are $3$ bonds leading to growth in site $(1,0)$, which results in the
configuration $i=1$, hence 
\begin{equation}
E_{1,3}={3 \over 3}x=0.2835.
\end{equation}
The upward growth process results in $i=2$ after truncation, and has 
probability
\begin{equation}
p_{\rm up}(3)=E_{2,3}={2 \over 3}(1+g_3(1)x)=0.7165.
\end{equation}
The third element in the column is $E_{3,3}=0$, which concludes the calculation
of the elements of the evolution matrix,
\begin{equation}
{\bf E}^{(3,1)}=\left[
\begin{array}{ccc}
0&0&0.2835 \\
1&0.4110&0.7165 \\
0&0.5890&0
\end{array}
\right],
\end{equation}
where the superscript indicates that it is the first-order approximation
for $N=3$.
The upward growth probabilities series is 
\begin{equation}
p_{\rm up}=(1,0.4110,0.7165),
\end{equation}
which happens to be equal to the second row of the matrix.

The normalized fixed point of the matrix is $P^*_1=0.0951$, $P^*_2=0.5695$
and $P^*_3=0.3354$. The average upward growth probability is
\begin{equation}
\langle p_{\rm up} \rangle^*=\sum_{i=1}^3P^*_ip_{\rm up}(i)=0.5695.
\end{equation}
We have performed some DLA simulations in the cylindrical geometry for 
several values of $N$ and measured $\langle p_{\rm up} \rangle^*$ 
\cite{Kol2000}.
The value obtained from simulations for $N=3$ is $0.5462$. The typical
accuracy is on the order of $10^{-4}$. 
The steady state average density and fractal dimension are evaluated using
Eqs. (\ref{rho}) and (\ref{estimate dimension}), 
\begin{eqnarray}
&&\rho={1\over 3\langle p_{\rm up} \rangle^*}=0.5853,~(0.6103), \nonumber \\
&&D=1-{\ln(\langle p_{\rm up} \rangle^*)\over \ln(3)}=1.5125,~(1.5506).
\end{eqnarray}
The values in parentheses are obtained from the same formulae, using the 
simulation value of $\langle p_{\rm up} \rangle^*$. 
The two other eigenvalues are complex,
$\lambda_{0,1}=-0.29\pm0.28i$, so according to Eq. (\ref{tau=}) $\tau=1.10$.

\subsection{Higher-order approximations for $N=3$}
\begin{figure}
\epsfbox{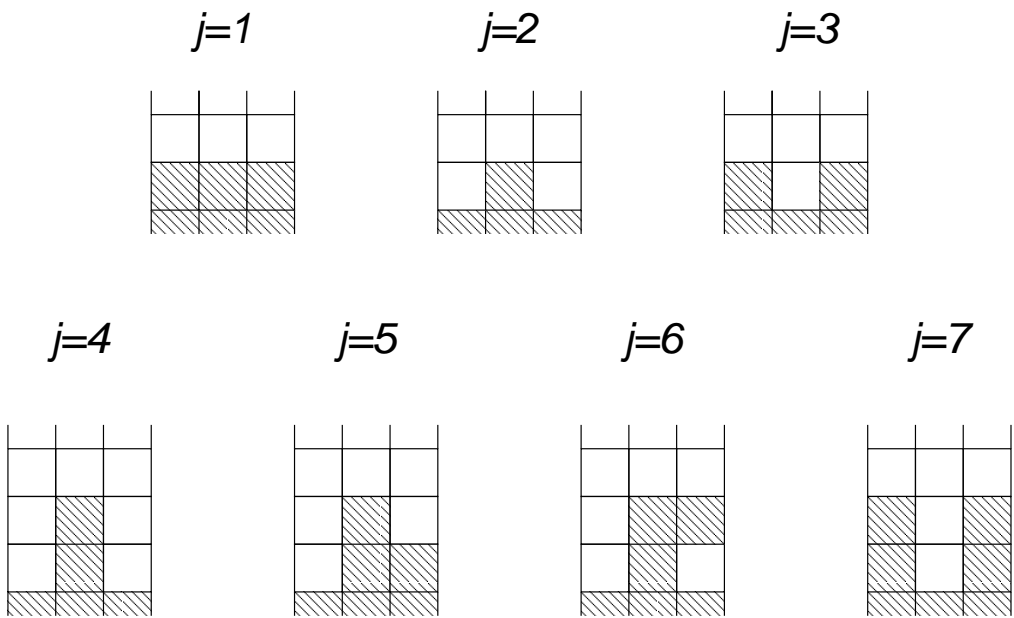}
\vspace{5mm}
\caption{Possible configurations in the second-order
approximation for $N=3$. Note that the first three configurations, $j=1,2,3$, 
are the same as in the first-order approximation.}
\label{N3o2index}
\end{figure}

The possible configurations of the interface in the second-order approximation
are listed and
indexed in Fig. \ref{N3o2index}. The growth probabilities for the first $3$ 
configurations were already computed in the previous section, but a 
rearrangement of the upward growths is required in the evolution matrix.
Now, the upward growth from configuration $j=2$ no longer stays at $i=2$, 
but rather makes a transition to $i=4$, and the upward growth from $j=3$ 
results in $i=5$ instead of $i=2$. Thus, we copy the previous
evolution matrix ${\bf E}^{(3,1)}$ into the upper left corner of the new
matrix ${\bf E}^{(3,2)}$ with the replacements: $E^{(3,2)}_{2,2}=0$,
$E^{(3,2)}_{4,2}=E^{(3,1)}_{2,2}$, $E^{(3,2)}_{2,3}=0$, and $E^{(3,2)}_{5,3}=
E^{(3,1)}_{2,3}$. The unspecified elements in the first three columns are all
equal to zero.

\begin{figure}
\epsfbox{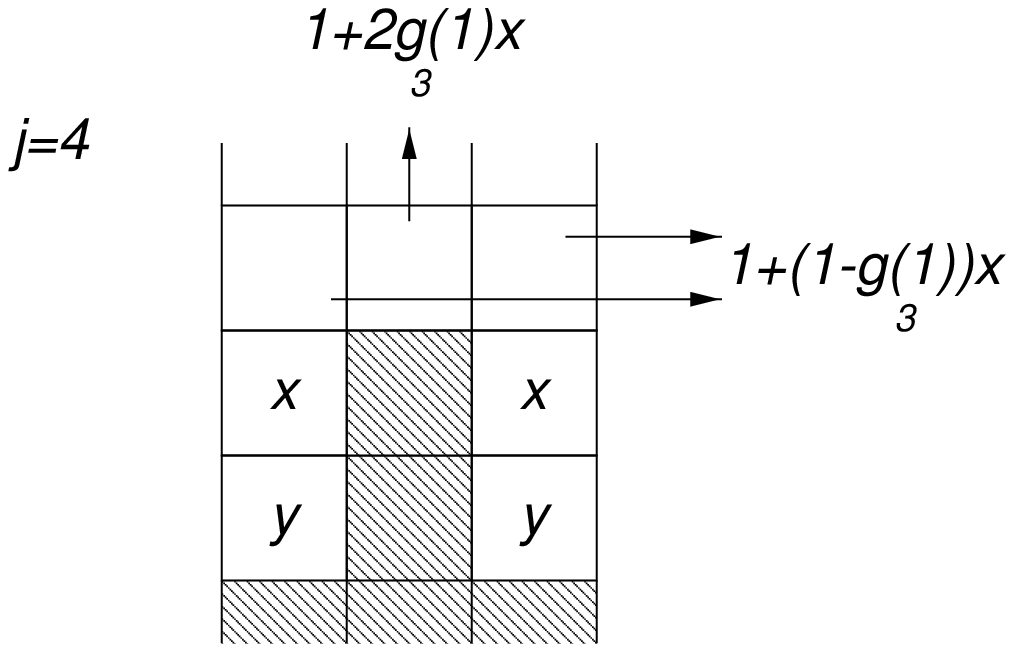}
\vspace{5mm}
\caption{The ``potential diagram'' for configuration $j=4$.}
\label{N3o2j=4}
\end{figure}

The next step is to go over each of the remaining configurations $i=4,\dots,7$,
and compute their probabilities, which are inserted into the 
evolution matrix according to the final configuration in which the relevant 
growth process results. Configuration $4$ is shown in Fig. \ref{N3o2j=4}. 
The Laplace equation is
\begin{eqnarray}
4y&=&y+x,\nonumber \\
4x&=&x+y+1+(1-g_3(1))x, \nonumber \\
\Rightarrow x&=&{3 \over 14}(7-\sqrt{21})=0.5180, \nonumber \\
y&=&x/3=0.1727.
\end{eqnarray}
The growth probabilities are
\begin{eqnarray}
E_{6,4}&=&{2 \over 3}x=0.3453, \nonumber \\
E_{5,4}&=&{4 \over 3}y={4 \over 9}x=0.2302, \nonumber \\
E_{4,4}&=&{1+2xg_3(1)\over 3}=0.4244.
\end{eqnarray}
The upward growth probability is $p_{\rm up}(4)=E_{4,4}=0.4244$.

\begin{figure}
\epsfbox{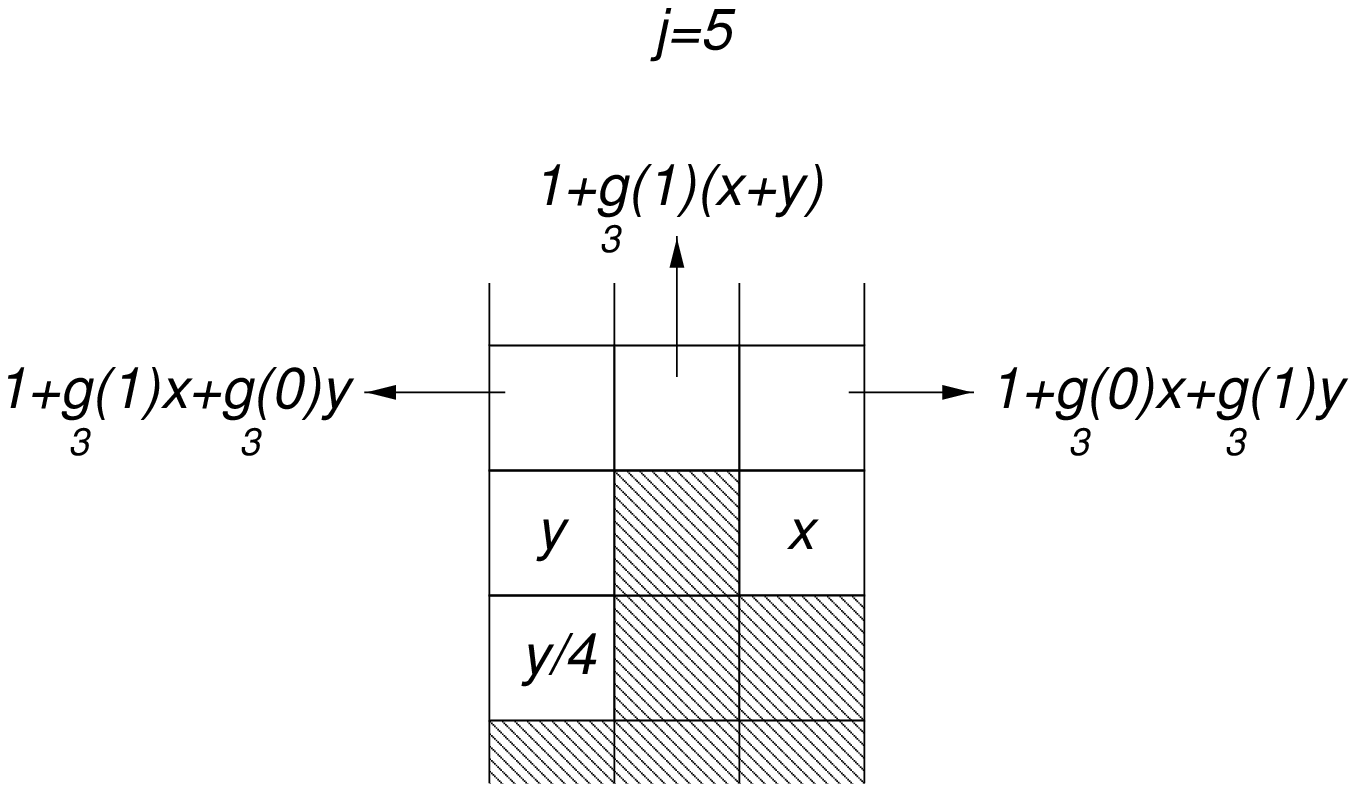}
\vspace{5mm}
\caption{The ``potential diagram'' for configuration $j=5$.}
\label{N3o2j=5}
\end{figure}

Configuration $5$ is shown in Fig. \ref{N3o2j=5}. The Laplace equations are
\begin{eqnarray}
&4y=y/4+x+xg_3(1)+yg_3(0)+1,& \nonumber \\
&4x=y+g_3(0)x+g_3(1)y+1,& \nonumber \\
&\Downarrow& \nonumber \\
&y=0.4808,& \nonumber \\
&x=0.4557.&
\end{eqnarray}
The growth probabilities are
\begin{eqnarray}
E_{7,5}&=&{2 \over 3}x=0.3038, \nonumber \\
E_{3,5}&=&{y \over 3}=0.1603, \nonumber \\
E_{2,5}&=&{3 \over 3}y/4=0.1202, \nonumber \\
E_{4,5}&=&{1+g_3(1)(x+y) \over 3}=0.4157. 
\end{eqnarray}
The upward growth probability is $p_{\rm up}(5)=E_{4,5}=0.4157$.

\begin{figure}
\epsfbox{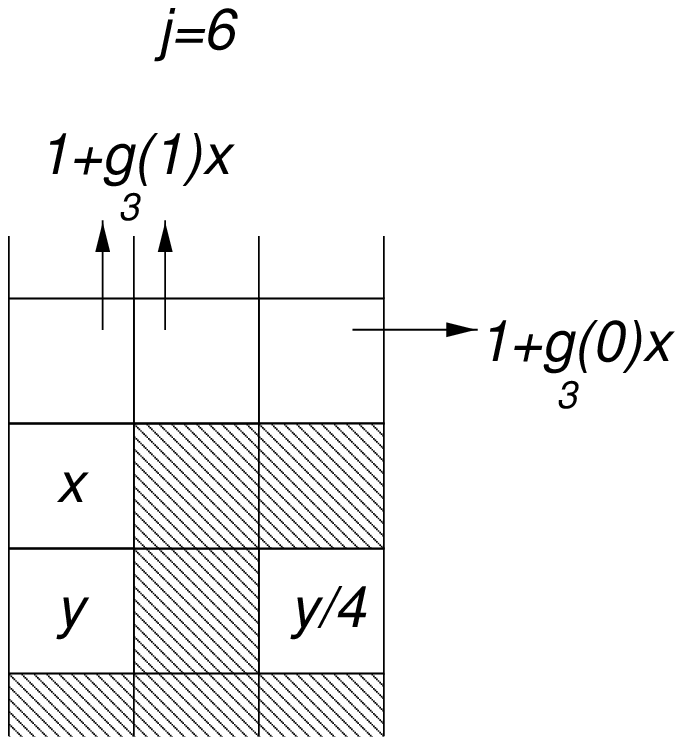}
\vspace{5mm}
\caption{The ``potential diagram'' for configuration $j=6$.}
\label{N3o2j=6}
\end{figure}

Configuration $6$ is shown in Fig. \ref{N3o2j=6}. The Laplace equations are
\begin{eqnarray}
&4y=y/4+x,& \nonumber \\
&4x=y+g_3(0)x+1,& \nonumber \\
&\Downarrow& \nonumber \\
&x={15 \over 151}(26-5\sqrt{21})=0.3067,& \nonumber \\
&y={4 \over 15}x=0.0818.&
\end{eqnarray}
The growth probabilities are
\begin{eqnarray}
E_{1,6}&=&{2 \over 3}x=0.2044, \nonumber \\
E_{3,6}&=&{2 \over 3}y={8 \over 45}x=0.0545, \nonumber \\
E_{7,6}&=&{3 \over 3}y/4=x/15=0.0204, \nonumber \\
E_{5,6}&=&{2 \over 3}(1+g_3(1)x)=0.7206.
\end{eqnarray}
The upward growth probability is $p_{\rm up}(6)=E_{5,6}=0.7206$.

\begin{figure}
\epsfbox{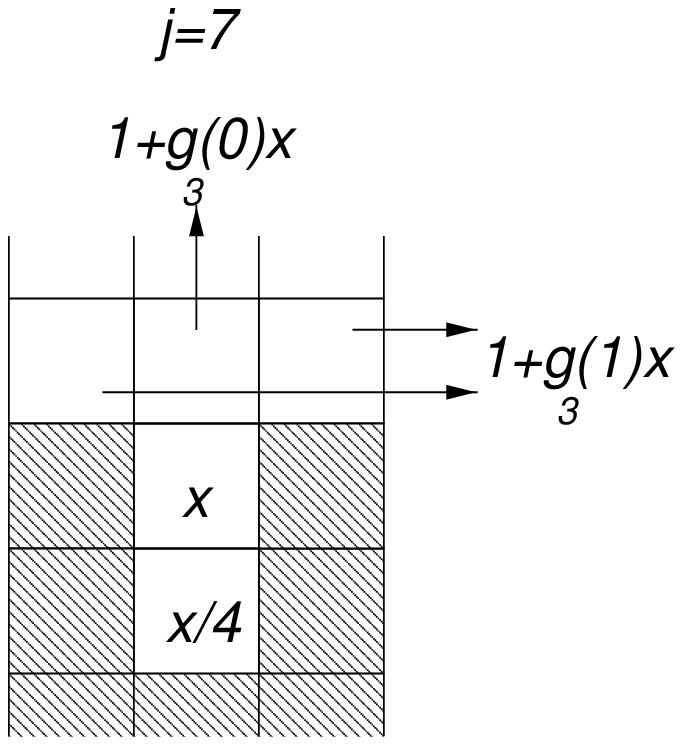}
\vspace{5mm}
\caption{The ``potential diagram'' for configuration $j=7$.}
\label{N3o2j=7}
\end{figure}

Configuration $7$ is shown in Fig. \ref{N3o2j=7}. The Laplace equations are
\begin{eqnarray}
&4x=x/4+g_3(0)x+1,& \nonumber \\
&\Downarrow& \nonumber \\
&x={12 \over 105}(21-4\sqrt{21})=0.3051.&
\end{eqnarray}
The growth probabilities are
\begin{eqnarray}
E_{1,7}&=&{2 \over 3}x=0.2304, \nonumber \\
E_{3,7}&=&{3 \over 3}x/4=0.0763, \nonumber \\
E_{5,7}&=&{2 \over 3}(1+g_3(1)x)=0.7203.
\end{eqnarray}
The upward growth probability is $p_{\rm up}(7)=E_{5,7}=0.7203$.

In summary,
\begin{eqnarray}
&&{\bf E}^{(3,2)}= \nonumber \\
&&\left[
\begin{array}{ccccccc}
0&0 &0.2835 &0 &0 &0.2044 &0.2034 \\
1&0 &0 &0 &0.1202 &0 &0 \\
0&0.5890 &0 &0 &0.1603 &0.0545 &0.0763 \\
0&0.4110 &0 &0.4244 &0.4157 &0 &0 \\
0&0 &0.7165 &0.2302 &0 &0.7206 &0.7203 \\
0&0 &0 &0.3453 &0 &0 &0 \\
0&0 &0 &0 &0.3038 &0.0204 &0
\end{array}
\right], \nonumber \\
\end{eqnarray}
\begin{eqnarray}
&&p_{\rm up}= \nonumber \\
&&\left(
\begin{array}{ccccccc}
1,&0.4110,&0.7165,&0.4244,&0.4157,&0.7206,&0.7203
\end{array}
\right). \nonumber \\
\end{eqnarray} 
One can check that elements in each column of the matrix sum up to $1$. 
Note that the majority of the elements are null.
The normalized fixed point is,
\begin{eqnarray}
{\bf P}^*=&\left( 
\begin{array}{cccc}
0.0685,&0.1011,&0.1145,&0.2680,
\end{array} \right.& \nonumber \\
&\left.
\begin{array}{ccc}
0.2711,&0.0925,&0.0843
\end{array}
\right)&
\end{eqnarray}
with which we compute some steady state quantities,
\begin{eqnarray}
&&\langle p_{\rm up} \rangle^*=\sum_{j=1}^7P^*_jp_{\rm up}(j)=0.5459,~(0.5462),
\nonumber \\
&&\rho={1 \over 3\langle p_{\rm up} \rangle^*}=0.6106,~(0.6103), \nonumber \\
&&D=1-{\ln(\langle p_{\rm up} \rangle^*)\over \ln(3)}=1.5510,~(1.5506),
\end{eqnarray}
where once again, the values from simulation are shown in parentheses. 
It is apparent that
the addition of configurations increases the accuracy of the results.
The eigenvalues with the largest absolute values (except for $1$) are
$\lambda_{0,1}=-0.34\pm0.40i$, hence $\tau=1.6$.
 
The third-order approximation yields $17$ configurations. The final
results are
\begin{eqnarray}
&&\langle p_{\rm up} \rangle^*=\sum_{j=1}^{17}P^*_jp_{\rm up}(j)=0.5460,
~(0.5462), \nonumber \\
&&\rho={1 \over 3\langle p_{\rm up} \rangle^*}=0.6104,~(0.6103), \nonumber \\
&&D=1-{\ln(\langle p_{\rm up} \rangle^*)\over \ln(3)}=1.5507,~(1.5506).
\end{eqnarray}
The eigenvalues with the largest absolute values (except for $1$) are
$\lambda_{0,1}=-0.34\pm0.40i$, hence $\tau=1.6$. 

\begin{figure}
\epsfbox{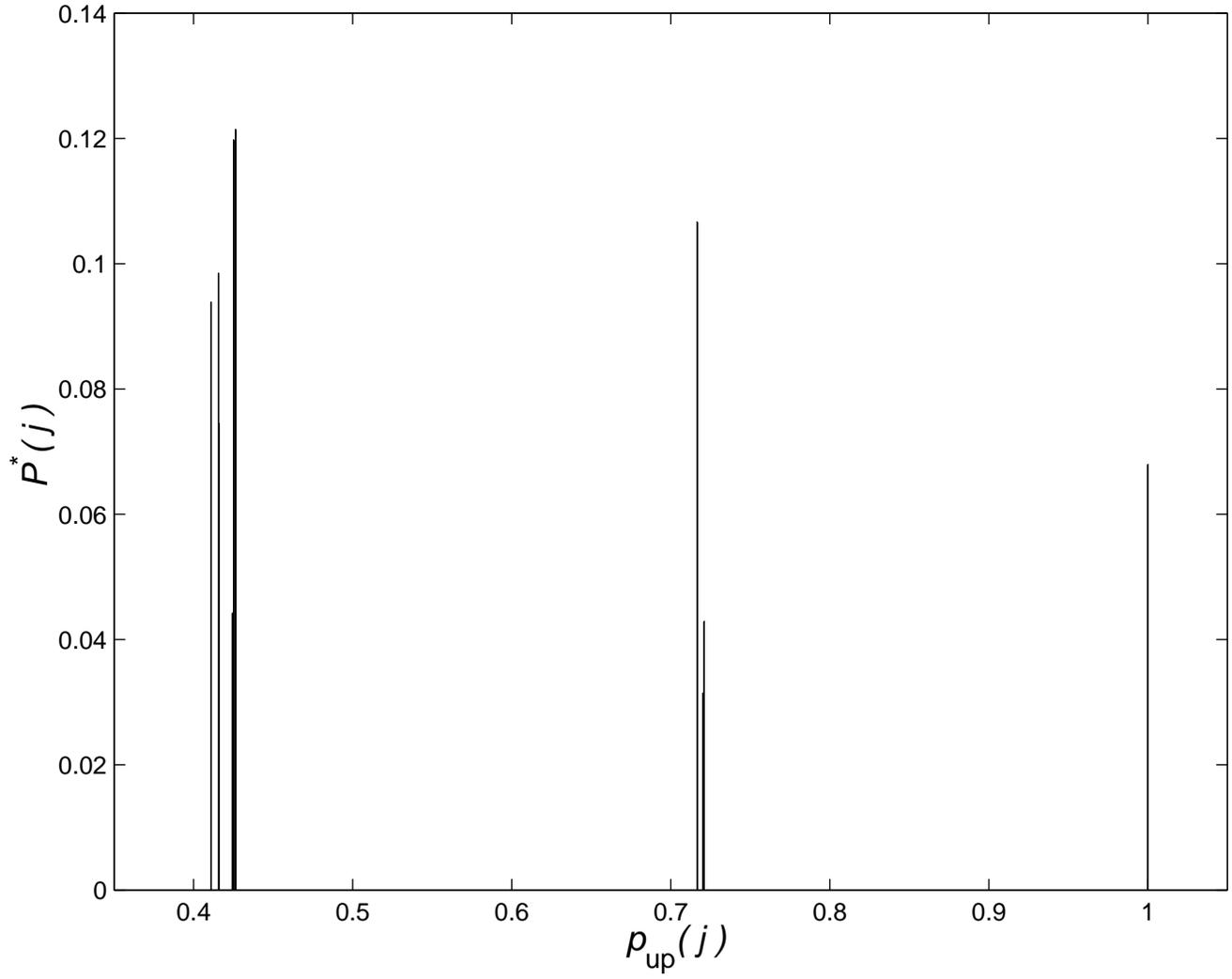}
\vspace{5mm}
\caption{The distribution of $p_{\rm up}$ over configurations for the 
third-order approximation for $N=3$.}
\label{pupdistribfig}
\end{figure}

It is interesting to inspect the histogram of
the distribution of $p_{\rm up}(j)$, illustrated in Fig. \ref{pupdistribfig}. 
One immediately observes that the upward growth 
probabilities are clustered in three groups: the top one at $1$, the second 
just above $0.7$ and the third, just above $0.4$. It is easy to check that the
top one corresponds to the configuration $i=1$, the middle group corresponds to
configurations that have two particles at the top row, and the bottom group
corresponds to configurations with one particle at the top row. This suggests,
that perhaps $17$ different configurations are excessive, and the real number
of effective configurations is around $3$. An interesting question is whether 
it is possible to further reduce the number of configurations in higher-order 
approximations by including only ``effective'' ones.

\subsection{First-order approximation for $N=4$}
Our last example is the case $N=4$, for which we present the first-order
calculation. First,
we calculate the Green's function $g_4(n)$ according to Eq. (\ref{g_N(n)}).
For $N=4$, there are four possible values for $k$ and $\kappa$, namely,
$k_l={2\pi \over N}l=0,{\pi \over 2},\pi,{3 \over 2}\pi$, $e^{-\kappa_0}=1$,
$e^{-\kappa_1}=e^{-\kappa_3}=2-\sqrt{3}$, and $e^{-\kappa_2}=3-\sqrt{8}$.
Hence,
\begin{eqnarray}
g_4(0)&=&{1+2(2-\sqrt{3})+3-\sqrt{8} \over 4} \nonumber \\
&=&2-{\sqrt{3}+\sqrt{2} \over 2}=0.4269, \nonumber \\
g_4(1)&=&g_4(3)={1-3+\sqrt{8} \over 4}={\sqrt{2}-1 \over 2}=0.2071, 
\nonumber \\
g_4(2)&=&{1-2(2-\sqrt{3})+3-\sqrt{8}\over 4}
={\sqrt{3}-\sqrt{2} \over 2}=0.1589. \nonumber \\
\end{eqnarray}
Once again, Eq. (\ref{norm}) is obeyed.

\begin{figure}
\epsfbox{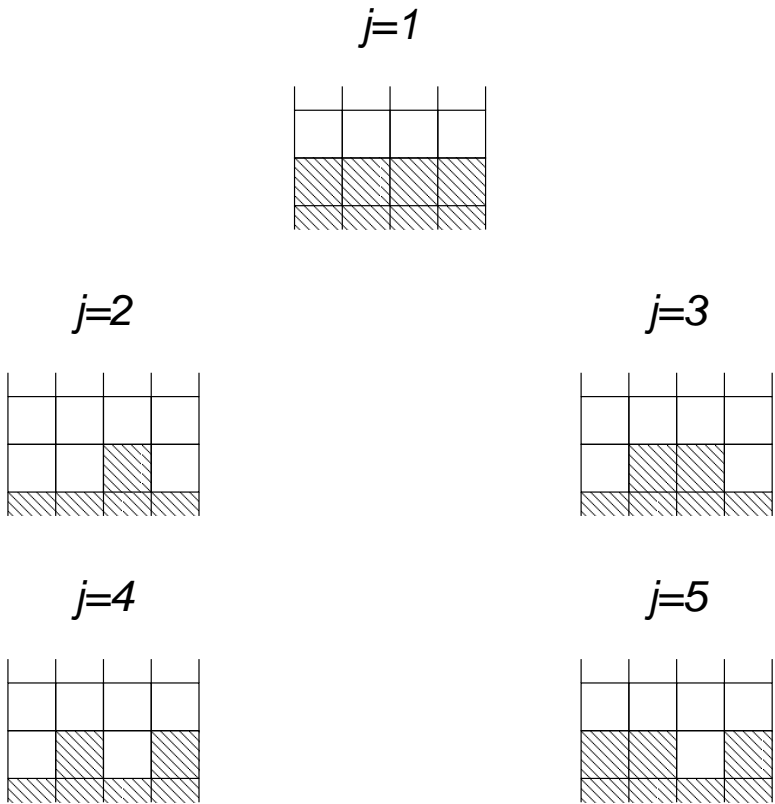}
\vspace{5mm}
\caption{Possible configurations in the first-order
approximation for $N=4$.}
\label{N4o1index}
\end{figure}

Figure \ref{N4o1index} displays the relevant configurations. 
Configuration 
$j=1$ grows into configuration $i=2$ with probability $1$, thus $E_{2,1}=1$ and
$E_{i,1}=0$ for $i\neq 2$. Also, $p_{\rm up}(1)=1$. 

\begin{figure}
\epsfbox{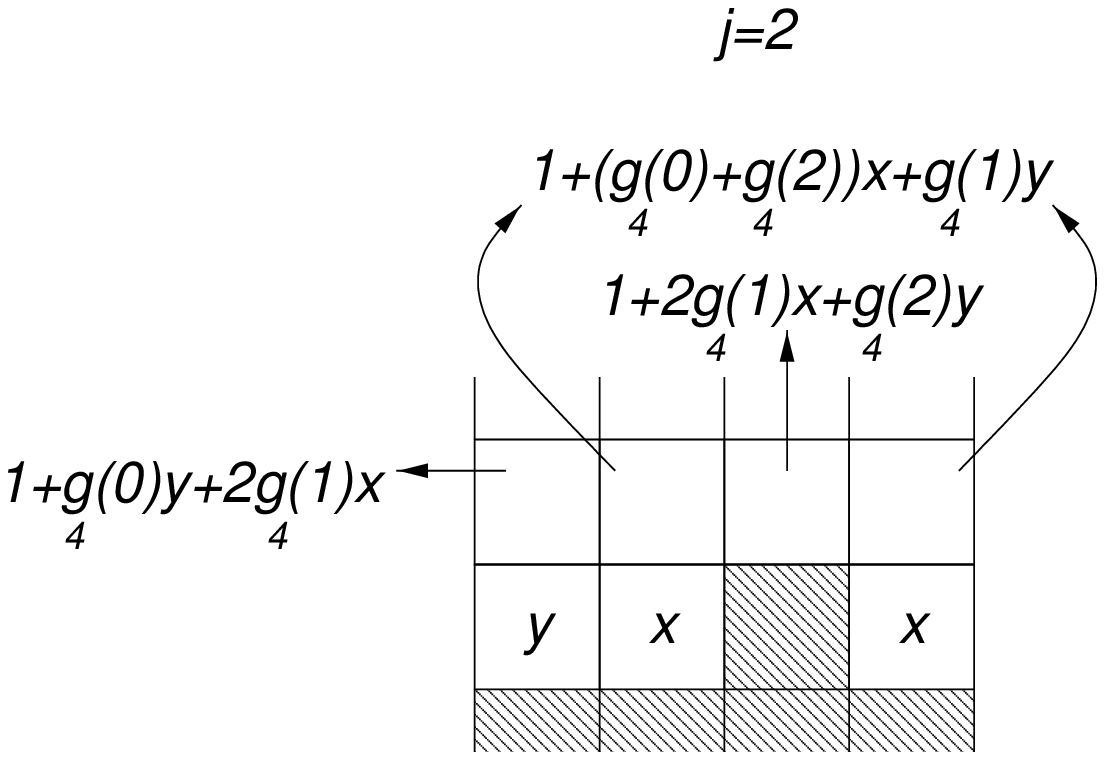}
\vspace{5mm}
\caption{The ``potential diagram'' for configuration $j=2$.}
\label{N4o1j=2}
\end{figure}

Configuration $j=2$ is shown in Fig. \ref{N4o1j=2}. The Laplace equations are
\begin{eqnarray}
&4x=y+g_4(1)y+\left(g_4\left(0\right)+g_4\left(2\right)\right)x+1,& 
\nonumber \\
&4y=2x+g_4(0)y+2g_4(1)x+1,& \nonumber \\
&\Downarrow & \nonumber \\
&x=0.5148,& \nonumber \\
&y=0.6277.&
\end{eqnarray}
The nonzero growth probabilities in the second column are
$E_{3,2}={4 \over 3}x=0.5148$, $E_{4,2}={1\over 4}y=0.1569$, and
$E_{2,2}={1 \over 4}\left(1+2g_4\left(1\right)x+g_4\left(2\right)y\right)
=0.3283=p_{\rm up}(2)$.

\begin{figure}
\epsfbox{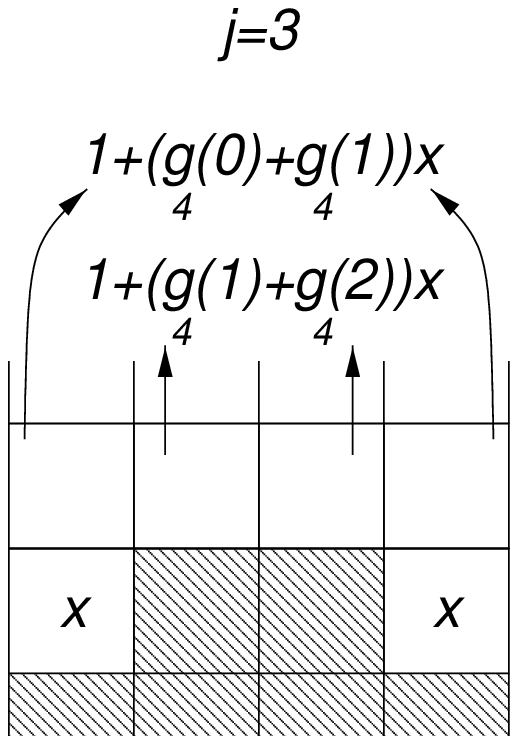}
\vspace{5mm}
\caption{The ``potential diagram'' for configuration $j=3$.}
\label{N4o1j=3}
\end{figure}

Configuration $j=3$ is presented in Fig. \ref{N4o1j=3}. The Laplace equation is
\begin{eqnarray}
4x&=&x+\left(g_4\left(0\right)+g_4\left(1\right)\right)x+1, \nonumber \\
\Rightarrow x&=&0.4226.
\end{eqnarray}
The nonzero growth probabilities in the third column are $E_{5,3}={4\over 4}x
=0.4226$ and $E_{2,3}={2\over4}\left[1+\left(g_4\left(1\right)+g_4\left(2
\right)\right)x\right]=0.5774=p_{\rm up}(3)$.

\begin{figure}
\epsfbox{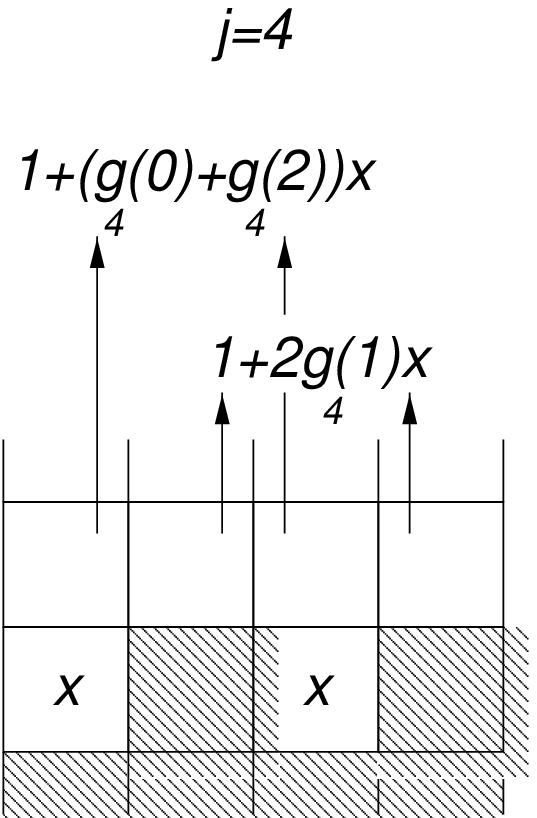}
\vspace{5mm}
\caption{The ``potential diagram'' for configuration $j=4$.}
\label{N4o1j=4}
\end{figure}

Configuration $j=4$ is shown in Fig. \ref{N4o1j=4}. The Laplace equation is
\begin{eqnarray}
4x&=&\left(g_4\left(0\right)+g_4\left(2\right)\right)x+1, \nonumber \\
\Rightarrow x&=&0.2929.
\end{eqnarray}
The nonzero growth probabilities in the fourth column are $E_{5,4}={6\over 4}
x=0.4393$ and $E_{2,4}={2\over 4}\left(1+2g_4\left(1\right)x\right)=0.5607
=p_{\rm up}(4)$. Note that this configuration already appeared for $N=2$.

\begin{figure}
\epsfbox{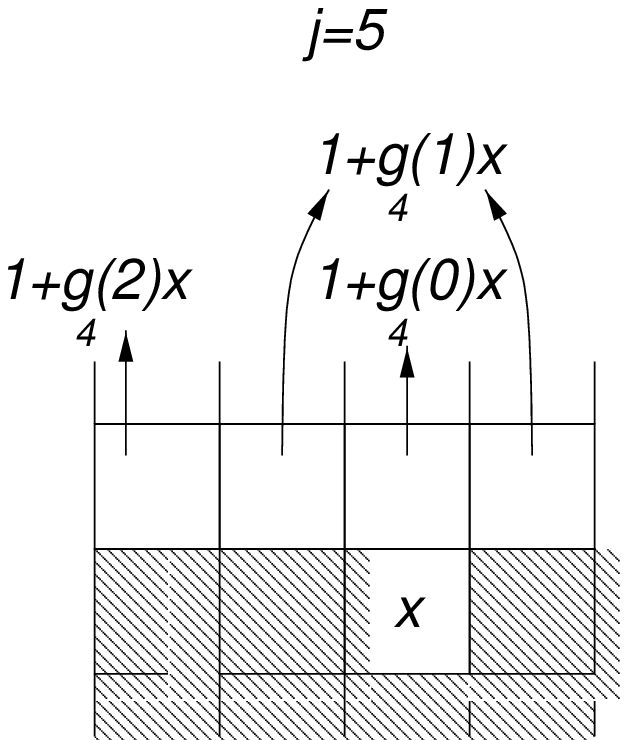}
\vspace{5mm}
\caption{The ``potential diagram'' for configuration $j=5$.}
\label{N4o1j=5}
\end{figure}

The last configuration is shown in Fig. \ref{N4o1j=5}. The Laplace equation is
\begin{eqnarray}
4x&=&1=g_4(0)x, \nonumber \\
\Rightarrow x&=&0.2799.
\end{eqnarray}
The nonzero growth probabilities in the fifth column are $E_{1,5}={3\over 4}x
=0.2099$ and $E_{2,5}={1\over 4}\left[3+\left(g_4\left(2\right)+2g_4\left(1
\right)\right)x\right]=0.7901=p_{\rm up}(5)$. This concludes the calculation
of the $5\times 5$ evolution matrix ${\bf E}^{(4,1)}$.

The steady-state vector is 
\begin{equation}
{\bf P}^*=\left(
\begin{array}{ccccc}
0.0298,&0.4954,&0.2551,&0.0777,&0.1420
\end{array}
\right).
\end{equation}
It enables to calculate the following steady-state quantities:
\begin{eqnarray}
&&\langle p_{\rm up}\rangle^*=P_2^*=0.4954,~(0.4657), \nonumber \\
&&\rho={1\over 4\langle p_{\rm up}\rangle^*}=0.5046,~(0.5368) \nonumber \\
&&D=1-{\ln(\langle p_{\rm up}\rangle^*)\over \ln(4)}=1.5066,~(1.5512),
\end{eqnarray}
where again, the values in parentheses are from simulation.
The eigenvalues with the largest absolute value after $1$ are $\lambda_{0,1}=
-0.16\pm0.38i$, hence $\tau=1.1$.

\begin{figure}
\epsfbox{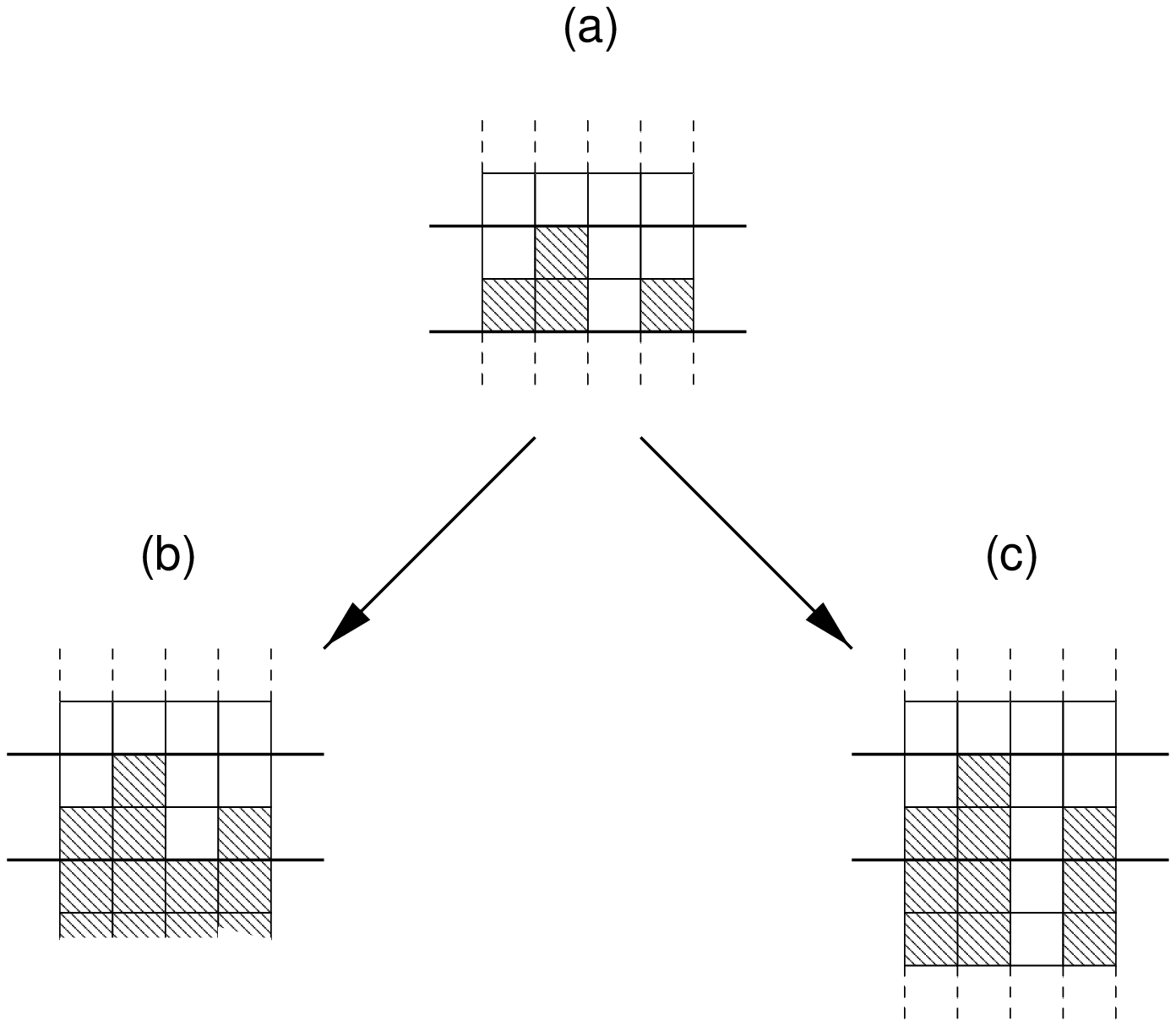}
\vspace{5mm}
\caption{The two top rows of a configuration are shown in (a). Two possible
extensions for the rest of the configuration below are (b), with
a filled
row right below the configuration (this boundary condition is used in the
calculations presented in this paper), or (c), with the bottom row of the 
configuration repeating itself ad infinitum, creating an infinite fjord.}
\label{bottombcfig}
\end{figure}

It is also possible to conduct these calculations using 
different boundary conditions at the bottom; rather than assuming that there 
is a filled row of occupied sites below the configuration, it
is possible to assume that each unoccupied site at the lowest row of the
configuration is above an infinite fjord that extends all the way below.
The two possibilities are explained in Fig. \ref{bottombcfig}. Performing the 
calculations with infinite fjords is a bit simpler, because there are less
configurations, e.g., the configuration $i=4$ would not appear in the 
first-order approximation for $N=4$ \cite{Kol2000}.

\subsection{Higher order computations}
As one increases $N$ and the order of approximation $O$, the number of 
configurations increases exponentially, and it becomes harder to go over all of
them manually. However, it is possible to construct a computer algorithm
to perform the procedure described here. The main challenges are the
automatic configuration recognition and automatic computation of the exact 
growth probabilities per configuration. 
In this section we explain the algorithm and report some of the important 
results.

The algorithm follows the method outlined in the examples of the 
previous sections, i.e., it goes over all the possible configurations
of the interface. In the sample calculations we have initially
made a list of all the possible configurations, called the index. 
Instead of doing this, the program 
starts with only one configuration, namely the flat one
(all the sites of the top row of the aggregate are occupied), which is indexed
by $j=1$. This configuration grows with probability $1$ to a new configuration
that has one particle at the top row, while the row below it is fully 
occupied. This new configuration is inserted into the list of 
configurations with an index $j=2$. Therefore, the program sets 
$E_{2,1}=1$ and $p_{\rm up}(1)=1$. Then the program continues by handling the
next configuration in the list, namely $j=2$. For each configuration, it 
solves the Laplace equations and calculates the growth probabilities. Each
growth process may create a new configuration. The resulting configuration is
first checked for consistency with the desired order $O$; configurations
which have $\Delta m>O$ are truncated, as in Fig. \ref{truncatefig}.
One then compares each
'new' configuration with
the existing list of configurations. If it does not exist in that list it 
is added at the end of the list, and indexed consecutively. If the index of the
configuration that results from the growth process is $i$ and the index of
the initial configuration is $j$ then the growth probability is inserted into
the matrix element $E_{i,j}$. The total sum of all the upward growth 
probabilities of the initial configuration $j$ is stored in $p_{\rm up}(j)$.
The main loop stops when the program finishes to process the last configuration
in the index list. At this stage the Markovian evolution matrix ${\bf E}$ is 
irreducible and closed, i.e., $\sum_iE_{i,j}=1$ for every $j$. Then
the fixed point ${\bf P}^*$ is calculated, by taking an
initial vector and iterating ${\bf E}$ on it many times until it converges 
(for very large matrices this is much faster than using any of the 
MATLAB library functions). 
The average upward growth probability is calculated using
\begin{equation}
\langle p_{\rm up} \rangle^*=\sum_jP^*_jp_{\rm up}(j),
\end {equation}
the average density and the fractal dimension are then computed using the left
hand side of Eq. (\ref{rho}) and Eq. (\ref{estimate dimension}).

\begin{figure}
\epsfbox{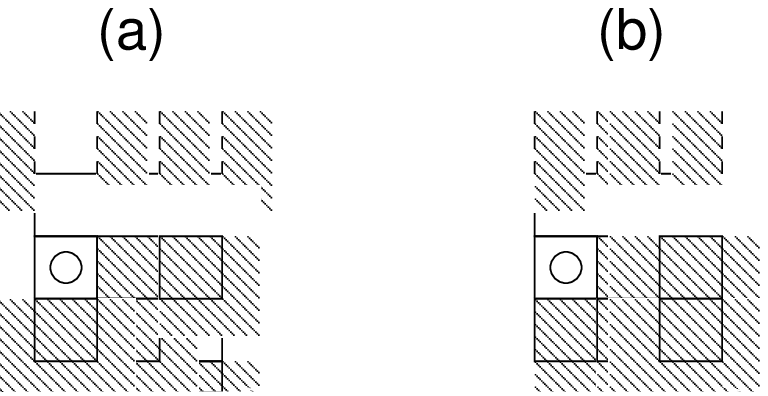}
\vspace{5mm}
\caption{Even though configuration (a) and (b) are not identical,
they are equivalent because they have the same growth probabilities.
Both configurations have the same external interface contour, which is
characterized by the set of sites that are connected to infinity. In this 
example there is only one such site, which is not higher that the aggregate,
and it is marked by a circle}
\label{N3example2fig}
\end{figure}

One of the challenges of the computer algorithm is the recognition of 
configurations. This recognition is important so that each growth process will
be inserted into the evolution matrix $E_{i,j}$ with the correct index $i$
($j$ is the index of the configuration before growth). The recognition
maybe difficult because configurations that seem different may actually be
equivalent. By equivalent we mean that they have the exact same set of 
transition (growth) probabilities. The solution to the Laplace equations 
is determined uniquely by the shape of the interface, therefore all of the
configurations with the same external interface are equivalent. 
The description of the interface is not a trivial task though. We find that an
efficient way to characterize an interface is by the set of empty sites that
are connected to infinity. Of course, it is sufficient to specify only empty
sites that are not higher than the highest particle in the aggregate, because
all of the empty sites above it are connected to infinity. Figure
\ref{N3example2fig} shows an example of two configurations that are not 
identical, but they have the same exterior contour. Both of them have a single
empty site that is connected to infinity.

In order to reduce greatly the number of configurations it is
advisable to take symmetry into account, i.e., all the configurations which can
be obtained from one another using a rotation around the axis of the cylinder 
have the same growth probabilities and the same steady state weights. The same
is true for mirror images. Instead of taking all of them into account, we 
choose one as a canonical representative of the whole set of symmetric
configurations.

The results are summarized in Table \ref{2dresults}. 
By comparing the approximations to accurate results from simulations, 
it seems that in order
to obtain a relative accuracy of about $10^{-3}$ one has to use at least 
an order of approximation of $O=N-2$ 
(except for $N=3$, where one still has to use the second-order approximation). 
This becomes very difficult already for $N=6$,
where in the fourth-order calculation there are $49678$ different 
configurations up to symmetry.

\section{Discussion}
\label{Discussionsec}
This paper treats DLA as a Markov process. The Markov states are the possible
shapes of the interface, and the Markovian evolution matrix ${\bf E}$
is calculated analytically using exact solutions of the 
Laplace equations, with proper normalizations. We propose
a truncation scheme that takes into account only a finite number of 
states. The states are ordered according to
the maximal difference in height between the highest and lowest points on
the interface, $\Delta m$, and in each order of truncation $O$, 
only the states with $\Delta m\leq O$ are included. 
We justify this approach by the fact that the
potential $\Phi$ decays exponentially in deep fjords, and thus the shape of
the interface in its deeper parts has very little effect on the growth 
probabilities. We perform this calculation for $N=2$, and verify that indeed 
it converges to the known analytic solution. We adopt the same approach for 
higher values of the width $N$, between $3$ and $7$, and calculate the average
density $\rho$ in good agreement with simulations. The fact that the
number of configurations grows exponentially with $N$ and with $O$, makes
the computation less effective than simulation for large $N$. 

We observe that the method converges as a function of $O$, 
also for higher values of $N$. 
Let us denote the calculated average steady-state density of an
aggregate of width $N$ in the $O$'th-order approximation by $\rho_c(N,O)$.
We observe that $\rho_c(N,O)$ converges to a finite limit very rapidly as
a function of $O$. In fact, a relative accuracy of $10^{-3}$ is achieved for 
$O=N-2$ (except for $N=3$). 
This enables us to obtain accurate results for $3 \leq N \leq 6$.
The drawback of this method is that the number of configurations diverges 
exponentially with $O$ and $N$, and therefore it is possible to perform
the calculations only for relatively low $N$'s and $O$'s.
Our computer was strong enough to perform the calculation only in the 
third-order approximation for $N=7$, and therefore the result for $N=7$ 
is not very accurate.
One would hope that it may be possible to perform low-order approximations for
large $N$'s and then extrapolate, in order to estimate the results for large 
$O$'s. Indeed, it is reasonable to conjecture 
the scaling law $\rho_c(N,O)=\rho(N)f(N/O)$, where $\rho(N)$ is the exact 
($O\to \infty$) density, as a function of $N$, and $f(N/O)$ is a universal
scaling function that obeys $\lim_{x\to 0}f(x)=1$. Our investigation shows
that in spite of the fact that the conjecture is not very accurate for $O=1$
and $O=2$, it is quite good for higher values of $O$, and presumably also for
higher values of $N$. This scaling relation may help to perform the 
extrapolation $O \to \infty$ for higher values of $N$. Paradoxically, it is 
very hard to obtain data points for large $N$'s and $O$'s, and thus to extract
the scaling function accurately. Thus we are unable to make the extrapolation
even for $N=7$, and we estimate $\rho(N)$ by the highest-order approximation
available. 
However, we suggest an alternative way to obtain 
$\rho_c(O,N)$, namely by simulation: it is possible to perform a regular
DLA simulation in cylindrical geometry, only that one has to keep the $O$'th 
row below the highest particle in the aggregate constantly filled. Measuring
the average density of the aggregate in such a simulation would approximate
$\rho_c(N,O)$. This simulation would be faster than a regular simulation, 
because particles would stick faster, due to the fact that they have less free
space. This study
would perhaps yield the scaling function $f(N/O)$, and enable extrapolation
of lower order approximations for higher $N$'s, should anyone venture to 
perform them on more powerful computers. In light of this discussion we 
suggest a more efficient way to perform DLA simulations in
cylindrical geometry. We argue that one can obtain a relative accuracy
of $10^{-3}$ if one follows just the $N-2$ top most rows of the aggregate.
This should save some time, because the diffusing particle would stick faster,
and it would also require less memory. This is not to say that it is sufficient
to grow the aggregate until it reaches a height of $N-2$, but rather, to 
perform many more growth processes, and each time the aggregate reaches
a height of $N-1$, truncate the bottom row.

We also discuss the temporal rate of convergence of the system to its
steady state. In this context we find that there is an
exponential convergence to the steady state, and we calculate the
characteristic time constant $\tau$. This is demonstrated using the simple
model of the frustrated climber.
The convergence is described in terms
of the eigenvalues of the Markovian matrix, and in terms of the infinite 
shift-down operator. 

Considering the fractal dimension, Pietronero {\em et al.} suggested 
that $\rho(N)=N^{D-d}$, as mentioned in Eq. (\ref{rho=N^(D-d)}). 
In principle, one should always include an amplitude
and finite size corrections of the form
\begin{equation}
\rho(N)=AN^{-\alpha}\left(1+B/N+\dots\right),
\label{rho=AN^}
\end{equation}
where $\alpha=d-D$, and $A$ and $B$ are constants.
The second term appearing in Eq. (\ref{rho=AN^}) is a correction to scaling
term. Generally, there is an infinite sum of such terms with higher 
negative powers of $N$. 
Because we have data only for small values of $N$, these correction terms
may be large, but since we have only a few accurate data points 
($\rho(N)$ for $N=2,3,\dots,6$), we try to extract the parameters 
$\alpha$, $A$ and $B$ only, and not higher order terms. 
Using the three results for
$N=4,5,6$, we determine the three unkown parameters to be
$A=0.82$, $B=0.35$
and $\alpha=0.362$, hence $D=1.64$.
The deviation from the well know value of $D=1.66$ can
be attributed to systematic error due to the omission of higher order 
finite size correction terms. We fit simulation data \cite{Kol2000} for 
$N=3$, $4$, $5$, $6$, $7$, $32$, $48$, $64$, $96$, $128$, to a higher-order
approximation $\rho(N)=AN^{-\alpha}\left(1+B/N+C/N^2\right)$, and find that
$C=-0.205$, $B=0.561$, $A=0.761$ and $\alpha=0.339$, which means that 
$D=1.661$. The maximum relative error of the fit is $1.2\times 10^{-3}$, and
the average relative error is $1.0 \times 10^{-3}$, which is in good
agreement with estimated accuracy of the simulations.

\acknowledgments
We wish to thank Barak Kol and A. Vespignani for helpful
discussions. We also wish to thank Yiftah Navot for helping with the
computer program, by suggesting more efficient data structures and algorithms.
We thank Nadav Schnerb for offering the frustrated climber
metaphor.
This work was supported by a grant from the German-Israeli Foundation (GIF).

\end{multicols}
\widetext

\begin{table}
\caption{The two-dimensional approximate 
results for various channel widths $N$ and for
different orders of approximation $O$. The quantities presented in each table
cell  are the average upward growth probability 
$\langle p_{\rm up} \rangle ^*$ and  the number of configurations $N_c$. 
The approximate results are compared with simulations.}
\begin{tabular}{c|c|c|c|c|c|c|c}
$N/O$&simulation&1&2&3&4&5&6\\
\tableline
3&0.5462&0.569489&0.545911&0.546046&0.546126&0.546132&0.546132 \\
&&3&7&17&45&127&371 \\
\tableline
4&0.4657&0.495435&0.464571&0.465395&0.465730&0.465765&0.465768 \\
&&5&20&98&575&3640&23676 \\
\tableline
5&0.4106&0.444088&0.407582&0.409497&0.410414&0.410547& \\
&&7&47&457&5539&69791& \\
\tableline
6&0.3696&0.405619&0.364352&0.367295&0.369172&& \\
&&12&131&2217&49678&& \\
\tableline
7&0.3377&0.375448&0.330112&0.333622&&& \\
&&17&337&10403&&& \\
\end{tabular}
\label{2dresults}
\end{table}

\begin{table}
\caption{Some steady state results of the third order approximation}
\begin{tabular}{c|ccccccc}
 &$\langle p_{\rm up} \rangle^*$&$P^*_0$&$P^*_1$&$P^*_2$&$P^*_3$&$P^*_4$
&$P^*_5$ \\
\tableline
$3$rd order&0.6812&0.2696&0.3114&0.1820&0.1029&0.0582&0.0329 \\
\tableline
Accurate&0.6812&0.2696&0.3113&0.1809&0.1032&0.0586&0.0332 \\
\end{tabular}
\label{steady3}
\end{table}


\begin{references}
\bibitem{remark}In \cite{Kol98} there are functions of the form
$a(j)=a(\infty)+x_0e^{-\alpha j}(1+x_1e^{-\beta j}+x_2e^{-2\beta j}+\dots)$,
where $a(\infty),x_0,\alpha>0,\beta>0,x_1,x_2,$ etc. are arbitrary constants.
For the present model, the average converges exponentially
$\langle a \rangle (t)=\langle a \rangle^*+p^t\sum_{j=0}^{\infty}a(j+t)v_j(0)
= \langle a \rangle^*+x_0p^te^{-\alpha t} \left[f_0+f_1e^{-\beta t}
+f_2e^{-2\beta t}+ \dots \right ]$,
where $\langle a \rangle^*\equiv \sum_{j=0}^{\infty}qp^ja(j)$, 
and
$f_i=x_i\sum_{j=0}^{\infty}e^{-(\alpha+i\beta)j}v_j(0)$.
Note that the
constant $a(\infty)$ does not affect the rate of convergence.
\end{references}
\end{document}